\documentclass{IEEEtran}

\usepackage{mathrsfs}
\usepackage{wrapfig}
\usepackage{picinpar}
\usepackage{times}
\usepackage{smallsec}

\usepackage[english]{babel}
\usepackage[latin1]{inputenc}
\usepackage{amsmath}
\usepackage{amssymb}

\usepackage{latexsym}

\usepackage{multicol}
\usepackage{hhline}
\usepackage{color}
\usepackage{xspace}

\newcommand{\ignore}[1]{}

\newcommand{\boxtheorem}{\hfill $\blacksquare$\vspace{2mm}}
\newcommand{\nit}[1]{{\it #1}}

\newcommand{\rred}[1]{\textcolor{red}{#1}}

\newcommand{\nn}{{\nit null}}

\newcommand{\n}{~{\it not}~}

\newcommand{\Vs}{V_{\!s}}
\newcommand{\cVs}{\mathcal{V}^{s}}
\newcommand{\ds}{D_{\!s}}

\newcommand{\bu}{\mathbf{bu}}
\newcommand{\au}{\mathbf{u}}
\newcommand{\s}{\mathbf{s}}
\newcommand{\bft}{\mathbf{t}}

\newcommand{\sql}{{\mbox{\tt \scriptsize sql}}\!}
\newcommand{\nulo}{{\mbox{\tt \scriptsize null}}\!}

\newcommand{\mc}[1]{\mathcal{ #1}}

\newcounter{theorem-counter}
\newcounter{corollary-counter}
\newcounter{lemma-counter}
\newcounter{definition-counter}
\newcounter{example-counter}
\newcounter{proposition-counter}
\newcounter{remark-counter}

{\vskip \abovedisplayskip \refstepcounter{theorem-counter}%
\noindent {\bf Theorem \arabic{theorem-counter}.}}%
{\boxtheorem}

{\vskip \abovedisplayskip \refstepcounter{corollary-counter}%
\noindent {\bf Corollary \arabic{corollary-counter}.}}%
{\boxtheorem}

{\vskip \abovedisplayskip \refstepcounter{lemma-counter}%
\noindent {\bf Lemma \arabic{lemma-counter}.}}%

\newenvironment{definition}%
{\vskip \abovedisplayskip \refstepcounter{definition-counter}%
\noindent {\bf Definition \arabic{definition-counter}.}}%
{\newline}

\newenvironment{example}%
{\vskip \abovedisplayskip \refstepcounter{example-counter}%
\noindent {\bf Example \arabic{example-counter}.}}%

\newenvironment{proposition}%
{\vskip \abovedisplayskip \refstepcounter{proposition-counter}%
\noindent {\bf Proposition \arabic{proposition-counter}.}}%

{\vskip \abovedisplayskip \refstepcounter{remark-counter}%
\noindent {\bf Remark \arabic{remark-counter}.}}%

\title{\bf
Achieving Data Privacy through Secrecy Views and Null-Based Virtual Updates}

\author{{\bf Leopoldo Bertossi}\thanks{Contact author: bertossi@scs.carleton.ca. \ Faculty Fellow of the IBM CAS, Toronto.} \ and \ {\bf Lechen Li}\\
 Carleton University,
School of Computer Science, Ottawa, Canada}

\begin{document}
\maketitle
\thispagestyle{empty}

\begin{abstract}
We may want to keep sensitive information in a relational database hidden from a user or group thereof. We
characterize sensitive data as the extensions of secrecy views.  The database,
before returning the answers to a query posed by a restricted user, is updated to make the secrecy views empty or a
single tuple with null values. Then,  a query  about any of those views returns no meaningful information. Since the
database is not supposed to be physically
changed for this purpose, the updates are only virtual, and also minimal. Minimality makes sure that
query answers, while being privacy preserving, are also maximally informative. The virtual updates are
based on null values as used in the SQL standard. We provide the semantics of secrecy views, virtual updates, and
secret answers to queries. The different instances resulting from the  virtually updates are specified as the models of
a logic  program with stable model semantics, which becomes the basis for computation of the secret answers.
\end{abstract}


\vspace{-4mm}
\section{Introduction} \label{sec:intro}
Database management systems allow for massive storage of data, which can be efficiently accessed and manipulated. However,
at the same time, the problems of data privacy are
becoming increasingly important and difficult to handle. For example, for commercial or legal reasons,
administrators of sensitive information may not want or be allowed to release
certain portions of the data. It becomes crucial to
address database privacy issues.

In this scenario,
certain users should have access to only certain portions of a database.
Preferably,
what a particular user (or class of them) is allowed or not allowed
to access should be specified in a declarative manner. This specification
should be used by the database engine when queries are processed and
answered. We would expect the database to return
answers that do not reveal anything that should be
kept protected from a particular user. On the other side and at the same time, the database should
return as informative answers as possible once the privacy conditions have been taken care of.

Some recent papers approach data privacy and access
control on the basis of {\em authorization views} \cite{RMS04,ZM05}. View-based data
privacy usually approaches the problem by specifying which views a
user {\em is allowed} to access. For example, when the database receives a query from
the user, it checks if the query can be answered using those views alone. More precisely, if the query can be
rewritten in terms of the views, for every possible instance \cite{RMS04}. If no {\em complete rewriting} is possible,
the query is rejected. In \cite{ZM05} the problem about the existence of a {\em conditional}
rewriting is investigated, i.e. relative to an instance at hand.

Our approach to the data protection problem is based on specifications of what users are {\em not} allowed to access
through query answers, which is quite natural. Data owners usually have a more clear picture of the data that
are sensitive rather than about the data that can be publicly released. Dealing with our problem as ``the complement"
of the problem formulated in terms of authorization views is not natural, and not necessarily easy, since complements
of database views would be involved \cite{jens02,jens03}.

According to our approach, the information to be protected is declared as a {\em secrecy view}, or a collection of them.
Their extensions have to be kept secret.
Each user or class of them may have associated a set of secrecy views. When a user poses a query to the database, the
system  virtually updates some of the attribute values on the basis of the secrecy views associated to that user.
In this work, we consider updates that modify attribute values through null values, which are commonly used to represent
missing or unknown values in incomplete databases. As
a consequence, in each of the resulting updated instances, the extension of each of the secrecy views either becomes empty
or contains a single tuple showing only null values. Either way, we say that {\em the secrecy view becomes null}. Then, the
original query is posed to the resulting class of updated instances. This amounts to: (a) Posing the query to each instance
in the class. (b) Answering it as usual from each of them. (c) Collecting the answers that are shared by all the instances
in the class. In this way, the system will return answers to the query that do not reveal the secret data.
The next example illustrates the gist of our approach.

\begin{example} \label{ex:intro} Consider the following relational database $D$:
{\small \begin{center}
 \begin{tabular}{c|c|c|c|}
  \hline
  $\nit{Marks}$ & $\nit{studentID}$ & $\nit{courseID}$ & $\nit{mark}$\\ \hline
   & $001$  & $01$& $56$\\
   & $001$  & $02$& $90$\\
   & $002$  & $02$& $70$\\
  \cline{2-4}
 \end{tabular}
\end{center}}

\noindent The \emph{secrecy view} $\Vs$ defined below specifies that a student with her course mark must be kept secret
when the mark is less than 60:

\centerline{$\Vs(\nit{sid}, \nit{cid}, \nit{mark}) \leftarrow \nit{Marks}(\nit{sid},\nit{cid},\nit{mark}), \nit{mark}< 60.\footnote{We use Datalog notation for view definitions, and sometimes also for queries.}$}

The view extension on the given instance is $\Vs(D) = \{\langle 001, 01, 56\rangle\}$, which is not null.
Now, a user  subject to this secrecy view wants to obtain the students' marks, posing the following query:
\begin{equation}\label{eq:q}
Q(\nit{sid},\nit{cid},\nit{mark}) \leftarrow \nit{Marks}(\nit{sid}, \nit{cid},\nit{mark}).
\end{equation}
Through this query the user can obtain the first record $\nit{Mark}(001, 01, 56)$, which is sensitive information.
A  way to solve this problem consists in \emph{virtually} updating the base relation according to the definition of the
secrecy view, making its extension null. In this way, the secret information, i.e. the extension of the secrecy view,
cannot be revealed to the user. Here, in order to protect the tuple $\nit{Mark}(001, 01, 56)$, the new instance $D'$
below is obtained by virtually updating the original instance, changing the attribute value $56$ into ${\tt NULL}$.

{\small
\begin{center}
\begin{tabular}{c|c|c|c|}
  \hline
  $\nit{Marks}$ & $\nit{studentID}$ & $\nit{courseID}$ & $\nit{mark}$\\ \hline
   & $001$  & $01$& ${\tt NULL}$\\
   & $001$  & $02$& $90$\\
   & $002$  & $02$& $70$\\
  \cline{2-4}
 \end{tabular}
\end{center}}

Now, by posing the query about the secrecy view, i.e.
\begin{eqnarray*}
Q_1(\nit{sid}, \nit{cid},\nit{mark}) &\leftarrow& \nit{Marks}(\nit{sid}, \nit{cid},\nit{mark}),\\&&\nit{mark} < 60,
 \end{eqnarray*}to  $D'$,
the user gets an empty answer, i.e. now
$\Vs(D') = \emptyset$. This is because -in SQL databases- the comparison of ${\tt NULL}$ with any other value is not
evaluated as true.

Now,
query (\ref{eq:q}) will get from $D'$ the first tuple with {\tt NULL} instead
of $56$, which can only be -misleadingly, expectedly and intendedly- interpreted by the user as
 an unknown or missing value for that student in the instance at hand $D$ (not $D'$, which is fully hidden to the user).
\boxtheorem
\end{example}
Notice that, among other elements (cf. end of Section \ref{sec:secretAns}), there are two that are crucial for this approach to work: (a) The given database may contain null values and if it has them or not is not known to the user, and (b) The
semantics of null values, including the logical operations with them. In this second regard, we can say for the moment and
in intuitive terms, that we will base our work on the SQL semantics of nulls, or,
more precisely, on a logical reconstruction of this semantics (cf. Sections \ref{sec:nvs} and \ref{app:nans}).

Hiding sensitive information is one of the concerns. Another one is about still providing as much information as possible
to the user. In consequence, the virtual updates have to be minimal in some sense, while still doing their job of protecting
data. In the previous example, we might consider virtually deleting the whole tuple $\nit{Marks}(001,05, 56)$ to protect
secret information, but we may lose some useful information, like the student ID and the course ID.
Furthermore, the user should not be able to guess the protected information by combing information obtained from different
queries.

As illustrated above, null values will be used to virtually update the database instance. Null values and incomplete databases
have received the
attention of the database community \cite{ZC82,VY79,IT84,LL99,ahv95}, and may have several possible interpretations, e.g. as a
replacement for a real value that is non-existent, missing, unknown, inapplicable, etc. Several formal semantics have been proposed for them. Furthermore, it is possible to consider different, coexisting null values. In this work, we will use a single null value, denoted as above and in the rest of this paper, by \nit{null}. Furthermore, we will treat \nit{null} as the {\tt NULL} in SQL relational databases.

We want our approach to be applicable to, and implementable on, DBMSs
that conform to the SQL Standard, and are used in database practice. We concentrate on that scenario and SQL nulls, leaving for possible future work the necessary modifications for our approach to work with other kinds of
null values.
Since the SQL standard does not provide a precise, formal semantics for {\tt NULL}, we define and adopt here a formal,
logical reconstruction of conjunctive query answering under SQL nulls (cf. Section \ref{app:nans}). In this direction,
we introduce unary predicates $\nit{IsNull}$ and $\nit{IsNotNull}$ in logical formulas that are true only when the argument
is, resp. is not, the constant  {\tt NULL}.
This treatment of null values was first outlined
in \cite{bb06}, but here we make it precise. It captures the logics and the semantics of the SQL {\tt NULL} that are
relevant for our work.\footnote{The main issue in \cite{bb06} was integrity constraint satisfaction in the presence of
nulls, for database repair and consistent query answering \cite{bertossi06}.} Including this aspect of nulls in our
work is necessary to provide the basic scientific foundations for our approach to privacy.

In this paper, we consider only conjunctive secrecy views and conjunctive queries. The semantics of null-based
virtual updates for data privacy that we provide is model-theoretic, in sense that the possible admissible instances
after the update, the so-called {\em secrecy instances}, are defined and characterized. This definition captures the
requirement that, on a secrecy instance, the extensions of the secrecy views contain only a tuple with null values or become
empty. Furthermore, the secrecy instances do not depart from the original instance by more than necessary to enforce secrecy.

Next, the semantics of {\em secret answers} to a query is introduced. Those answers are invariant under the class of secrecy
instances. More precisely, a ground tuple $\bar{t}$ to a first order query $\mc{Q}(\bar{x})$ is a secret answer from
instance $D$ if it is an answer to $\mc{Q}(\bar{x})$ in every possible secrecy instance for $D$. Of course, explicitly
computing and materializing all the secrecy instances to secretely answer a query is too costly. Ways around this naive
approach have to be found.

Actually, we show that the class of secrecy instances, for a given instance $D$ and set of secrecy views $\mathcal{V}^s$,
can be captured in terms of a disjunctive logic program with stable model semantics  \cite{gelfond91,LGaij}. More
precisely, there is a one-to-one correspondence between the secrecy instances and the stable models of the program.
 As a consequence, the logic
programs can be used to: (a) Compactly specify (axiomatize) the class of secrecy instances; and (b) Compute secret answers
to queries by running the program on top of the original instance.

Our work has some similarities with that on {\em database repairs} and {\em consistent query answering} (CQA)
\cite{bertossi06,bertossi11}. In that case, the problem is about restoring consistency of a database wrt to a set
of integrity constrains by means of minimal updates. The alternative consistent instances that emerge in this way are
called {\em repairs}. They can be used to characterize the consistent data in an inconsistent database as the one
that is invariant under the class of repairs. It is possible to specify the repairs of a database by means of disjunctive
logic programs with stable model semantics (cf. \cite{bertossi11} for references on CQA).

Summarizing, in this paper we make the following contributions: (a) We introduce {\em secrecy views} to specify what to hide from a given
user. (b) We introduce the virtual {\em secrecy instances} that are obtained by minimally changing attribute values by nulls,
to make the secrecy view extensions null. (c) We introduce the {\em secret answers} as those that are certain for the class
of secrecy instances. Those are the answers returned to the user. (d) We establish that this approach works in the sense that
the queries about the secrecy view contents always return meaningless answers; and furthermore, the user cannot reconstruct
the original instance via secret answers to different queries. (e) We provide a precise logical characterization of query
answering in databases with null values {\em \`a la} SQL. (f) We specify by means of logic programs the secrecy instances of
a  database, which allows for skeptical reasoning, and then, certain query answering, directly from the specification. (g)
We establish sme connections between secret query answering and CQA in databases.

The structure of the rest of this paper is as follows. In Section \ref{sec:preli} we introduce
basic notation and definitions, including the semantics of conjunctive query
answering in databases with nulls.
In Section \ref{sec:secInstance}, we introduce the secrecy instances and investigate the properties of secrecy.
Section \ref{sec:secretAns} presents the notion of secret answer to a query. Section \ref{sec:logicProm} presents secrecy
logic programs. Section \ref{sec:cqa} investigates the connection to database repairs and consistent query answering.
Section \ref{sec:related} discusses related work. In Section \ref{sec:conclusions} we draw conclusions, and point to
future work.

\vspace{-2mm}
\section{Preliminaries}\label{sec:preli}

Consider a relational schema $\Sigma=(\mc{U},\mc{R},\mc{B})$,
where $\mc{U}$ is the possibly infinite database domain, with $\nn \in \mc{U}$, $\mc{R}$ is a finite set of
database predicates,  and $\mc{B}$ is a finite set of built-in predicates, say $\mc{B} = \{=,\neq, >,<\}$. For an $n$-ary predicate $R \in \mc{R}$, $R[i]$ denotes the $i$th position or attribute of $R$, with $1 \leq i \leq n$. The schema determines a
language $L(\Sigma)$ of first-order (FO) predicate logic,
with predicates in $\mc{R} \cup \mc{B}$ and {\em constants} in $\mc{U}$. A relational
{\em instance} $D$ for schema $\Sigma$ is a finite set of ground atoms of the form $R(\bar{a})$, with $R \in \mc{R}$, and $\bar{a}$
a tuple of constants from $\mc{U}$ \cite{ahv95}.

A query is a formula $\mc{Q}(\bar{x})$ of $L(\Sigma)$, with $n$ free variables $\bar{x}$. $D \models \mc{Q}[\bar{c}]$
denotes
that instance $D$ makes $\mc{Q}$ true with the free variables taking values as in $\bar{c}$ $\in$ $\mc{U}^n$.
In this case, $\bar{c}$ is an answer to the query. $\mc{Q}(D)$ denotes the set of answers to query $\mc{Q}$
from $D$. We will concentrate on {\em conjunctive queries}, that are $L(\Sigma)$-formulas
 consisting of a possibly empty prefix of existential quantifiers followed by a conjunction of (database or
built-in) atoms.

\begin{example}\label{ex:first} Consider the following database instance $D_1$:
{\small \begin{center}
\begin{tabular}{c|c|c|}\hline
  $R$ & $A$ & $B$ \\ \hline
   & $a$ & $b$\\
   & $c$ &  $d$\\
   & $e$ &  $\nn$\\
  \cline{2-3}
 \end{tabular}
~~~
 \begin{tabular}{c|c|c|}\hline
  $S$ & $B$ & $C$ \\ \hline
   & $b$ & $f$ \\
   & $d$ & $g$\\
   & $\nn$ & $j$ \\
  \cline{2-3}
 \end{tabular}
\end{center} }
\noindent For the conjunctive query  $\mc{Q}_1(x,z)\!: \exists y (R(x,y) \wedge S(y,z))$, it holds, e.g. \ $D_1 \models \mc{Q}_1[a,f]$.
Actually, $\mc{Q}_1(D_1) = \{\langle a, f\rangle,$ $\langle c, g \rangle, \langle e, j\rangle\}$. Notice that here, and for the moment, we are
treating $\nn$ as any other constant in the domain.
\boxtheorem
\end{example}
Data will be protected via a fixed set ${\cal V}^s$ of secrecy views $\Vs$. They are associated to a particular user or class of them.

\begin{definition}
A \emph{secrecy view} $\Vs$ is defined by a Datalog rule of the form
\begin{equation} \Vs(\bar x) \leftarrow R_1(\bar{x}_1), \ldots,
R_n(\bar{x}_n), \ \varphi, \label{eq:view}
\end{equation} with $R_i \in \mathcal{R}, \bar{x} \subseteq \bigcup_i \bar{x}_i$ and
$\bar{x}_i$ is a tuple of  variables.\footnote{We will frequently use Datalog notation for
view definitions and queries. When there is no possible
confusion, we treat sequences of variables as set of variables. I.e. $x_1 \cdots x_n$ as $\{x_1, \ldots, x_n\}$.}
Formula $\varphi$ is a conjunction of built-in atoms containing {\em terms}, i.e. domain constants or variables. \ignore{(each of the latter appearing in some of the $R_i(\bar{x}_i)$).}
\boxtheorem
\end{definition}
We can see that a secrecy view is defined by a conjunctive query with built-in predicates written in $L(\Sigma)$.
The conjunctive query associated to the view in (\ref{eq:view}) is:
\begin{equation}
\mc{Q}^{\!\Vs}\!(\bar{x}): \ \exists \bar{y}(R_1(\bar{x}_1) \wedge \cdots \wedge
R_n(\bar{x}_n) \wedge \varphi), \label{eq:conj}
\end{equation} with $\bar{y} = (\bigcup \bar{x}_i) \smallsetminus \bar{x}$.  \ $\nit{Conj}(\Sigma)$ denotes the class
of conjunctive queries of $L(\Sigma)$, and
$\Vs(D)$ the extension of view $\Vs$ computed on instance $D$ for $\Sigma$. By definition,
$\Vs(D) = \mc{Q}^{\!\Vs}(D)$.

\ignore{In Section \ref{app:nans}
we provide a precise definition of conjunctive query with built-ins as needed for our work and applications
to privacy.}

\begin{example}\label{ex:svdef} (example \ref{ex:first} cont.) For the given instance $D_1$, consider  the
secrecy view defined by~ $\Vs(x) \leftarrow R(x,y),S(y,z)$. Here, the
data protected by the view are those that belongs to its extension, namely $\Vs(D_1) = \{\langle a\rangle, \langle c\rangle, \langle e \rangle\}$. Sometimes, to
emphasize the view predicate involved,  we write instead $\Vs(D_1) = \{\Vs(a), \Vs(c), \Vs(e)\}$. The corresponding conjunctive query is \ $\mc{Q}^{\!\Vs}(x)\!: \ \exists y \exists z(R(x,y) \wedge S(y,z))$.
\boxtheorem
\end{example}
 Finally, an {\em integrity constraint} (IC) is a sentence $\psi$ of $L(\Sigma)$. $D \models \psi$ denotes that instance
 $D$ satisfies $\psi$. For a fixed set $\mc{I}$ of ICs, we say that $D$ is {\em consistent} when $D \models \mc{I}$, i.e.
when $D$ satisfies each element of $\mc{I}$.

For both of the notions of query answer and IC satisfaction above we are using  the classic concept of satisfaction
 of predicate logic, denoted with $\models$. According to it, the constant $\nn$ is treated as any other constant of the database domain. We will use this
notion at some places. However, in order to capture the special role of $\nn$ among those constants, as in SQL databases, we will introduce next a different
notion, denoted with $\models_{\!_N}$. In Example \ref{ex:first}, under the new semantics, and due to the participation
of $\nn$ in join,  the tuple $\langle e, j \rangle$ will not be an answer anymore, i.e. $D_1 \not \models_{\!_N} \mc{Q}_1[e,j]$. The two notions, $\models$ and $\models_{\!_N}$, will coexist and also be related
(cf. Section \ref{app:nans}).

\vspace{-2mm}
\subsection{Null value semantics: The gist} \label{sec:nvs} \vspace{-1mm}
In \cite{codd79}, Codd proposed a three-valued logic with truth values \emph{true}, \emph{false}, and \emph{unknown} for
relational databases with {\tt NULL}. When a {\tt NULL} is involved in a comparison operation, the result is \emph{unknown}.
This logic has been adopted by the SQL standard, and partially implemented
in most common commercial DBMSs (with some variations). As a result, the semantics of
{\tt NULL} in both the SQL standard and the
commercial DBMSs is not quite clear; in particular, for IC satisfaction in the presence of {\tt NULL}.

The semantics for IC satisfaction with {\tt NULL}
 introduced in \cite{bb06, bravo07} presents a FO semantics for nulls in SQL databases. It is a reconstruction in classical logic of the treatment of {\tt NULL}  in
SQL DBs. More precisely, this semantics captures the notion of satisfaction of ICs, and also of query answering for a broad
class of queries in relational databases.  In the rest of this section, we motivate and sketch some of the elements of the notion of
query answer that we will use  in the rest of this work. The details can be found in Section  \ref{app:nans}. In the
following, we assume that there is a single constant, \nit{null}, to represent a null value.

A tuple $\bar{c}$ of elements of $\mc{U}$ is an answer to query $\mc{Q}(\bar{x})$, denoted $D \models_N \mc{Q}(\bar{c})$,
if the formula (that represents) $\mc{Q}$ is {\em classically true} when the
quantifiers on its {\em relevant} variables (attributes) run over $({\cal U} \smallsetminus \{\nn\})$; and
those on of the non-relevant variables run over ${\cal U}$. The free relevant variables cannot  take the value $\nn$
either. \ignore{In the classical setting, $\nn$ is treated as any other constant.}  For a
precise definition  see Section \ref{app:nans} (and also \cite{bb06,bravo07}).

\begin{example} \label{ex:ansquery} Consider  the instance $D_2$ and query below:

{\small \begin{center}
\begin{tabular}{c|c|c|c|}
  \hline
  $R$ & $A$ & $B$ & $C$ \\ \hline
   & $1$ & $1$ & $1$\\
   & $2$ &  $\nn$ & $\nn$\\
   & $\nn$ &  $3$ & $3$\\
  \cline{2-4}
 \end{tabular}
~~~
 \begin{tabular}{c|c|}
  \hline
  $S$ & $B$  \\ \hline
   & $\nn$  \\
   & $1$ \\
   & $3$ \\
  \cline{2-2}
 \end{tabular}
\end{center}}

\begin{equation}
\mc{Q}_2(x):
 \ \exists y \exists z(R(x,y,z)\wedge S(y) \wedge y>2). \label{eq:quer}
 \end{equation}
 A variable $v$ (quantified or not) in a conjunctive query is {\em relevant} if it appears (non-trivially) twice
 in the formula after the quantifier prefix \cite{bb06}. Occurrences of the form $v = \nn$ and $v \neq \nn$ do not count though.
 In query (\ref{eq:quer}),  the only relevant quantified variable is $y$, because it participates in a join and a built-in in the
 quantifier-free matrix of (\ref{eq:quer}).  So, there are two reasons for $y$ to be relevant. The only free variable is $x$, which is not relevant.
  As for query answers, the only candidate values for $x$ are: $\nn, 2, 1$. In this case,
 $\nn$ is a candidate value because $x$ is a non-relevant variable.

First, $x
= \nn$
is an answer to the query, because the formula $\exists y \exists z
(R(x,y,z)\wedge S(y) \wedge y > 2)$ is true in $D_2$, with a non-null witness value for $y$
and a witness value for $z$ that combined make the (non-quantified) formula true. Namely, $y=3, z=3$. So, it holds $D_2 \models_N \mc{Q}_2[\nn]$.

Next, $x = 2$ is not an answer. For this value of $x$, because the candidate value for $y$, namely $\nn$ that accompanies $2$ in $P$, makes the formula $(R(x,y,z)\wedge S(y)\wedge y > 2)$ false. Even if it were true, this value for $y$ would not be allowed.

Finally, $x = 1$ is not an answer, because the only candidate value for $y$, namely $1$, makes the formula false. In consequence, $\nn$ is the only answer.
\boxtheorem
\end{example}
This notion of query answer coincides with the classic FO semantics for queries and databases without null
values \cite{bb06,bravo07}. The next example with SQL queries and {\tt NULL} provides
additional intuition and  motivation for the formal semantics of Section
\ref{app:nans}. Notice the use in logical queries of the new unary predicates \nit{IsNull} and \nit{IsNotNull}  that we also formally introduce
 in Section \ref{app:nans}.

 \begin{example} \label{ex:nulls} Consider the schema $\mc{S} = \{R(A,B)\}$ and the instance in the table below. In it $\tt{NULL}$ is the SQL null. If this instance is stored in an SQL
 database, we can observe the behavior of the following queries when they are directly translated
 into SQL  and run on  an SQL DB:

 \vspace{-2mm}
 \begin{multicols}{2}
\begin{tabular}{c|c|c|} \hline
      $R$  &  $A$  &  $B$ \\ \hline
       &   a &   b \\
         & a   &  c\\
         & d &   $\tt{NULL}$\\
        &  d &   e\\
        &  u  &  u\\
        &  v  &  $\tt{NULL}$\\
     &     v  &   r \\
     & $\tt{NULL}$ & $\tt{NULL}$ \\ \hhline{~--}
     \end{tabular}

     \begin{tabular}{c|c|c|} \hline
      $S$  &  $B$  &  $C$ \\ \hline
    &b  &  h\\
    &{\tt NULL} & s\\
    &l &   m\\ \hhline{~--}
    \end{tabular}

\hspace*{-12mm}\noindent (a)  $\mc{Q}_1(x,y)\!: R(x,y) \ \wedge \ y = \nn$

\hspace*{-1cm}SQL:\hspace{1mm}\verb+Select * from R+
\hspace*{0.2cm} \verb+where B = NULL;+

\hspace*{-1cm}Result: No tuple
    \end{multicols}
\vspace{-2mm}\noindent (b) \
 $\mc{Q}_1'(x,y)\!: R(x,y) \ \wedge \ \nit{IsNull}(y)$

SQL: Now uses {\tt IS}\hspace{1mm}{\tt NULL}

Result:  $\langle \mbox{d}, {\tt NULL}\rangle$, $\langle \mbox{v}, {\tt NULL}\rangle$, $\langle {\tt NULL}, {\tt NULL}\rangle$

\noindent (c) \ $\mc{Q}_2(x,y): \ R(x,y)  \wedge  y \neq \nn$

SQL: \verb+Select * from R where B <> NULL;+

Result: No tuple

\noindent (d) \ $\mc{Q}_2'(x,y): \ R(x,y)  \wedge  \nit{IsNotNull}(y)$

SQL: Now uses {\tt IS}\hspace{1mm}{\tt NOT}\hspace{1mm}{\tt NULL}

Answer: The five expected tuples

\noindent (e) \
$\mc{Q}_3(x,y): R(x,y) \wedge x = y$

SQL: \verb+Select * from R where A = B;+

Result: \ $\langle \mbox{u},\mbox{u}\rangle$

\noindent (f) \
$\mc{Q}_4(x,y): R(x,y) \wedge  x \neq y$

SQL: \verb+Select * from R where A <> B;+

Result: Four tuples: \ $\langle \mbox{a},\mbox{b}\rangle, \langle \mbox{a},\mbox{c}\rangle, \langle \mbox{d},\mbox{e}\rangle, \langle \mbox{v},\mbox{r}\rangle$

\noindent (g) \
$\mc{Q}_5(x,y,x,z):  R(x,y) \wedge R(x,z) \wedge y \neq z$

SQL: \verb+Select * from R r1, R r2 where+

      \verb+      r1.A = r2.A and r1.B <> r2.B;+

Result: \ $\langle \mbox{a},\mbox{b},\mbox{a},\mbox{c}\rangle, \langle \mbox{a},\mbox{c},\mbox{a},\mbox{b}\rangle$

\noindent (h) \ $\mc{Q}_6(x,y,z,t)\!: R(x,y) \wedge S(z,t) \wedge y=z$

SQL: \ \hspace{-1mm}\verb+Select * from R r1, S s1+\\
\hspace*{15mm}\verb+where r1.B = s1.B;+

Result: $\langle \mbox{a},\mbox{b},\mbox{b},\mbox{h}\rangle$

\noindent (i) \ SQL: \ \verb+Select * from R r1 join S s1+\\
\hspace*{3cm} \verb+on r1.B = s1.B;+

Result:\footnote{The same result is obtained from
DBMSs that do not require an explicitly equality together with the join.} \ $\langle \mbox{a},\mbox{b},\mbox{b},\mbox{h}\rangle$

\noindent (j) $\mc{Q}_7(x,y,z,t): R(x,y) \wedge S(z,t) \wedge y \neq z$

SQL: \ \verb+Select R1.A, R1.B, S1.B, S1.C+\\
 \hspace*{0.7cm}\verb+from R R1, S S1 where R1.B <> S1.B';+

Result: $\langle\mbox{a},\mbox{c},\mbox{b},\mbox{h}\rangle, \langle \mbox{d},\mbox{e},\mbox{b},\mbox{h}\rangle, \langle \mbox{u},\mbox{u},\mbox{b},\mbox{h}\rangle, \langle \mbox{v},\mbox{r},\mbox{b},\mbox{h}\rangle,$\\
\hspace*{1mm}$\langle \mbox{a},\mbox{b},\mbox{l},\mbox{m}\rangle, \langle \mbox{a},\mbox{c},\mbox{l},\mbox{m}\rangle, \langle \mbox{d},\mbox{e},\mbox{l},\mbox{m}\rangle,$ $\langle \mbox{u},\mbox{u},\mbox{l},\mbox{m}\rangle,\langle
\mbox{v},\mbox{r},\mbox{l},\mbox{m}\rangle$
\boxtheorem
\end{example}

\vspace{-4mm}
\subsection{Semantics of query answers with nulls}\label{app:nans} \vspace{-1mm}
Here we  introduce the semantics of FO conjunctive query answering in relational databases with null values.\footnote{This semantics can be extended to a broader class of queries and also to integrity constraint satisfaction. It builds upon a
similar and more general semantics first introduced
in  \cite{bb06,bravo07}.}
More precisely, in SQL relational databases with a single null value, $\nit{null}$, that is handled like the SQL {\tt NULL}. The  SQL queries are first reconstructed as queries in the FO language $L(\Sigma^\nulo)$ associated to $\Sigma^\nulo = (\mc{U}, \mc{R}, \mc{B}^\nulo)$, with $\mc{B}^\nulo
= \mc{B} \cup \{{\it IsNull}(\cdot), {\it IsNotNull}(\cdot)\}$. The last two are new
unary built-in predicates that correspond to the SQL predicates
{\tt IS}\hspace{1mm}{\tt NULL} and {\tt IS}\hspace{1mm}{\tt NOT}\hspace{1mm}{\tt NULL}, used to check
null values.
Their intended semantics is as follows (cf. Definition \ref{df:nqs}): ${\it IsNull}(\nn)$ is true, but ${\it IsNull}(c)$ is
false for any other constant $c$ in the database domain. And, for any constant $d \in \mc{U}$, ${\it IsNotNull}(d)$ is
true iff ${\it IsNull}(d)$ is false.

Introducing these predicates is necessary, because, as shown in Example \ref{ex:nulls}, in the presence of {\tt NULL}, SQL treats {\tt IS}\hspace{1mm}{\tt NULL} and
{\tt IS}\hspace{1mm}{\tt NOT}\hspace{1mm}{\tt NULL}  differently from $=$ and $\neq$, resp.  For example, the queries $\mc{Q}(x)\!:  \exists y(R(x,y)\wedge\nit{IsNull}(y))$ and $\mc{Q}'(x)\!:$ $\exists y(R(x,y)\wedge y = \nn)$
 are both conjunctive queries of $L(\Sigma^\nulo)$, but in
 SQL relational databases, they have different semantics.

In Example \ref{ex:nulls}, each query $\mc{Q}$ is
defined by the formula $\psi$ on the right-hand side. Below, we will identify the query with its defining FO formula.
Furthermore, we exclude from the SQL-like conjunctive queries those like (a) and (c) in Example \ref{ex:nulls}.
\begin{definition} (a) The class $\nit{Conj}^\sql(\Sigma^\nulo)$ contains all the {\em conjunctive queries} in
$L(\Sigma^\nulo)$ of the form
\begin{equation}
\mc{Q}(\bar{x}): \ \exists \bar{y}(A_1(\bar{x}_1) \wedge \cdots \wedge A_n(\bar{x}_n)), \label{eq:conjunctive}
\end{equation}
where $\bar{y} \subseteq \bigcup_i \bar{x}_i$, $\bar{x} = (\bigcup_i \bar{x}_i) \smallsetminus \bar{y}$, and
the $A_i$ are atoms containing any of the predicates in $\mc{R} \cup \mc{B}^\nulo$
plus terms, i.e. variables or constants in $\mc{U}$. Furthermore, those atoms are never
of the form  $t = \nn, \ \nn = t, \ t \neq \nn, \nn \neq t$, with $t$ a term, $\nn$ or  not. \\
(b) With $\nit{Conj}(\Sigma^\nulo)$ we denote the class of all conjunctive queries of the form (\ref{eq:conjunctive}),
but without the restrictions on (in)equality atoms imposed on $\nit{Conj}^\sql(\Sigma^\nulo)$. \boxtheorem
\end{definition}
The idea here is to force conjunctive queries {\em \`a la} SQL, i.e. those in  $\nit{Conj}^\sql(\Sigma^\nulo)$, that
explicitly mention the null value in (in)equalities, to use the built-ins $\nit{InNull}$
or $\nit{IsNotNull}$.  Notice that the class $\nit{Conj}(\Sigma^\nulo)$
includes both $\nit{Conj}^\sql(\Sigma^\nulo)$ and $\nit{Conj}(\Sigma)$.

\begin{definition}
Consider  a query in $\nit{Conj}(\Sigma^\nulo)$  of the form
$\mathcal{Q}(\bar{x})\!\!: \exists \bar{y} \psi(\bar{x},\bar{y})$, with $\exists \bar{y}$ a possibly empty prefix of existential quantifiers, and $\psi$
is a quantifier-free conjunction of atoms.
A variable $v$ is  {\em relevant} for $\mathcal{Q}$ \cite{bravo07} if it occurs at least twice in  $\psi$, without
considering the atoms ${\it IsNull}(v),$ ${\it IsNotNull}(v),$ $v~\theta~\nn,$ or $\nn~\theta~v,$ with
$\theta \in \mc{B}$. $\mc{V}^R(\mathcal{Q})$ denotes the set of relevant variables for $\mc{Q}$.
\boxtheorem
\end{definition}
For example, for the query $\mc{Q}(x): \exists y(P(x,y,z)\wedge Q(y)\wedge\nit{IsNull}(y))$, $\mc{V}^R(\mc{Q}(x)) = \{y \}$, because $y$ is used twice in the subformula $P(x,y,z)\wedge Q(y)$. \ignore{Actually, the notion of relevant variable can be applied to any first-order formula written in prenex normal form.}

As usual in FO logic, we consider assignments from the set, $\nit{Var}$, of variables to the underlying database domain $\mc{U}$ (that contains
constant $\nn$), i.e. $s: \nit{Var} \rightarrow
\mc{U}$. Such an assignment can be extended to terms, as $\bar{s}$. It maps every variable $x$ to $s(x)$, and every element $c$ of
$\mc{U}$ to $c$.  For an assignment $s$, a variable $y$ and a constant $c$, $s\frac{y}{c}$ denotes the
assignment that coincides with $s$ everywhere, possibly except on $y$, that takes the value $c$. \ignore{We define similarly the extension $\bar{s}\frac{y}{c}$.} Given a formula $\psi$, $\psi[s]$ denotes the formula obtained from $\psi$ by replacing its free variables by their values according to $s$.

Now, given a formula (query) $\chi$ and a variable assignment function $s$, we verify if instance $D$ satisfies $\chi[s]$ by assuming that the quantifiers on relevant variables range over $({\cal U} \smallsetminus \{\nn\})$, and
those on non-relevant variables range over ${\cal U}$. More precisely, we define, {\em by induction on} $\chi$, when
$D$ satisfies $\chi$ with assignment $s$, denoted $D \models_{\!_N}
\chi[s]$.

\begin{definition} \label{df:nqs} Let $\chi$ be a
query in $\nit{Conj}(\Sigma^\nulo)$,
and $s$ an
assignment.  The pair $D,s$ satisfies $\chi$ under the null-semantics,
denoted $D \models_{\!_N} \chi[s]$, exactly  in
the following cases: \ (below $t,t_1, \ldots$ are terms; and $x, x_1,x_2$  variables)

\noindent     1. (a) $D \models_{\!_N} \nit{IsNull}(t)[s]$, with $s(t) =
\nn$. (b) $D \models_{\!_N} \nit{IsNotNull}(t)[s]$, with $s(t) \neq
\nn$.

\noindent 2. $D \models_{\!_N} (t_1 < t_2)[s]$, with $\bar{s}(t_1) \neq \nn \neq \bar{s}(t_2)$, and $\bar{s}(t_1) < \bar{s}(t_2)$ \ (similarly for $>$).\footnote{Of course, when there is an order relation on $\mc{U}$.}

     \noindent 3.
      (a) $D \models_{\!_N} (x = c)[s]$, with $s(x) = c \in
(\mc{U}\smallsetminus \{\nn\})$.  (or
symmetrically).\footnote{Here we
use the symbols $=$ and $\neq$ both at the object and the meta levels, but
there should not be a confusion since valuations are involved.}

     \noindent  (b) $D \models_{\!_N} (x_1 = x_2)[s]$, with $s(x_1) = s(x_2)
\neq \nn$.

\noindent  (c) $D \models_{\!_N} (c = c)[s]$, with $c \in (\mc{U}\smallsetminus
\{\nn\})$.

\noindent 4.  (a) $D \models_{\!_N} (x \neq c)[s]$, with $\nn \neq s(x) \neq c \in (\mc{U}\smallsetminus
\{\nn\})$.
(or symmetrically).

\noindent (b) $D \models_{\!_N} (c_1 \neq c_2)[s]$, with $c_1 \neq c_2$, and $c_1, c _2 \in (\mc{U}\smallsetminus
\{\nn\})$.

\noindent
5. $D \models_{\!_N} R(t_1,\dots,t_n)[s]$, with $R \in {\cal R}$, and $R(\bar{s}(t_1),$ $\dots,$ $\bar{s}(t_n)) \in D$.

\noindent  6. $D \models_{\!_N} (\alpha \wedge \beta)[s]$, with $\alpha$, $\beta$ quantifier-free, $s(y) \neq \nn$ for
every $y \in \mc{V}^R(\alpha \wedge \beta)$, and
  $D \models_{\!_N} \alpha[s]$ and $D \models_{\!_N} \beta[s]$.

\noindent  7. $D \models_{\!_N} (\exists y \ \alpha)[s]$ when: (a) if $y \in {\cal V}^R(\alpha)$, there is $c$ in $({\cal U} \smallsetminus \{\nn\})$ with $D \models_{\!_N} \alpha[s \frac{y}{c}]$; or (b) if $y \not \in {\cal V}^R(\alpha)$, there is $c$ in ${\cal
    U}$ with $D \models_{\!_N} \alpha[s \frac{y}{c}]$.\boxtheorem
\end{definition}
This semantics can be applied  to conjunctive queries in $\nit{Conj}^\sql(\Sigma^\nulo)$.
The notion of relevant attribute and this semantics of query satisfaction can be both extended to more complex
formulas. In particular, they can be applied also to the satisfaction of integrity constraints under SQL null values
\cite{bravo07,bb06}.

\begin{definition}~\cite{bravo07} \label{def:nvs}  Let $\mathcal{Q}(\bar{x}): \exists \bar{y} \psi(\bar{x},\bar{y})$ be
a conjunctive query in $\nit{Conj}(\Sigma^\nulo)$, with $\bar{x} = x_1, \ldots, x_n$.
\ignore{ \ (a) {\bf OLD. TO DELETE.} A tuple $(t_1,\dots,t_n) \in \mathcal{U}^n$
is an {\em answer from $D$ under the null query answering
semantics} to $\mc{Q}$, in short, an $N${\em -answer}, and denoted $D \models_{\!_N} \mathcal{Q}[t_1,\ldots,t_n]$, iff there exists an assignment $s$ such
that: i. $s(x_i)=t_i$, for $i=1,\dots,n$; ii. $t_i \neq \nn$, for each $x_i \in \mc{V}^R(\mathcal{Q})$;
and iii. $D
\models_{\!_N} (\exists \bar{y} \psi)[\bar{s}]$.  }

\noindent (a)  A tuple $\langle c_1,\dots,c_n\rangle \in \mathcal{U}^n$
is an {\em answer from $D$ under the null query answering
semantics} to $\mc{Q}$, in short, an $N${\em -answer}, denoted $D \models_{\!_N} \mathcal{Q}[c_1,\ldots,c_n]$, iff there exists an assignment $s$ such
that  $s(x_i)=c_i$, for $i=1,\dots,n$; and $D
\models_{\!_N} (\exists \bar{y} \psi)[s]$.

\noindent (b) $\mc{Q}^N\!(D)$ denotes the set of
$N$-answers to $\mc{Q}$ from instance $D$. Similarly,  $V^{\!N\!}(D)$ denotes a view extension
according to the $N$-answer semantics: $V^{\!N\!}(D) = (\mc{Q}^V\!){\!^N\!}(D)$.

\noindent (c) If $\mc{Q}$ is a sentence (boolean query), the $N$-answer  is {\it yes} iff $D \models_{\!_N} \mc{Q}$,
and $\nit{no}$, otherwise. \boxtheorem
\end{definition}
Notice that $D
\models_{\!_N} (\exists \bar{y} \psi)[s]$ in (a) above requires, according to Definition \ref{df:nqs}, that the
variables in the existential prefix $\exists \bar{y}$ that are  relevant do not take
the value $\nn$. The free variables $x_i$ in $\mc{Q}(\bar{x})$ may take the value $\nn$ only when they are not
 relevant in the query. Example \ref{ex:ansquery} illustrates this definition. In it, since the free variable $x$ is not relevant,
 $\mc{Q}_2^N\!(D_2) =\{\langle \nn\rangle\}$. Similarly, in Example \ref{ex:first}, it holds:
 $\mc{Q}_1^N\!(D_1) = \{\langle a, f\rangle,$ $\langle c, g \rangle\} \subseteq \mc{Q}_1(D_1)$.

 Actually, it is easy to prove that, for queries in $\nit{Conj}(\Sigma^\nulo)$, it holds in general:
 $\mc{Q}^N\!(D) \subseteq \mc{Q}(D)$.
 Furthermore, the $N$-query answering semantics coincides with classical FO query answering semantics in databases without null values \cite{bravo07,bb06}. More precisely, if $\nn \notin \mc{U}$ (and then it does not appear in $D$ or $\mc{Q}$ either): \ $D \models_{\!_N} \mc{Q}[\bar{t}]$ \  iff  \ $D \models \mc{Q}[\bar{t}]$.

 Furthermore,
every  conjunctive query in $\nit{Conj}(\Sigma^\nulo)$ can be syntactically transformed into a new FO query for which
the evaluation can be done by treating {\em null} as any other constant \cite{bravo07,bb06}. (A similar transformation will
be found in Proposition \ref{prop:admis} below.)

More precisely, a conjunctive query $\mc{Q}(\bar{x}) \in \nit{Conj}(\Sigma^\nulo)$, i.e. of the form (\ref{eq:conjunctive}),  can be
rewritten into a classic conjunctive query, as follows:
\begin{equation}
\mc{Q}^\nit{rw\!}(\bar{x}): \exists \bar y(A_1(\bar{x}_1) \wedge \cdots \wedge A_n(\bar{x}_n) \ \wedge
\! \! \! \!\bigwedge_{v\in \mc{V}^R(\mc{Q})} \!\!\! v \neq \nn). \label{eq:rew}
\end{equation}
 It holds: \ $D \models_{\!_N} \mc{Q}[\bar{c}]  \ \ \mbox{ iff } \  \ D \models \mc{Q}^\nit{rw\!}[\bar{c}]$. Here, on the right-hand side, we have classic FO satisfaction, and $\nn$ is treated as an ordinary constant in the domain. This transformation ensures that  relevant variables range over $({\cal U} \smallsetminus \{\nn\})$. Query $\mc{Q}^\nit{rw\!}(\bar{x})$ belongs to $\nit{Conj}(\Sigma^\nulo)$, and it may contain
atoms of the form  $\nit{IsNull}(t)$ or $\nit{IsNotNull}(t)$. However, replacing them by $t = \nn$ or $t \neq \nn$, resp.,
leads to a query in $\nit{Conj}(\Sigma)$ that has the same answers as (\ref{eq:rew}) (under the same classic semantics).
\begin{example} (example \ref{ex:ansquery} continued) Query $\mc{Q}$ in (\ref{eq:quer}) can be rewritten as

\vspace{1mm}
\centerline{$\mc{Q}_2^\nit{rw\!}: \ \exists y\exists z
(P(x,y,z)\wedge Q(y)\wedge y > 2\wedge y\neq\nn).$}

\vspace{1mm}\noindent
  We had $D \not \models_{\!_N} \mc{Q}_2[1]$. Now also $D \not \models \exists y\exists z(P(1,y,z)\wedge Q(y)\wedge y>2\wedge y\neq\nn)$
 under classic query evaluation, with $\nn$ treated as an ordinary constant.
 Similarly, $D \not \models \mc{Q}_2^\nit{rw\!}[2]$ due to the new
 conjunct $y=\nn$.
Finally, $D \models \mc{Q}_2^\nit{rw\!}[\nn]$ because $D \models (P(\nn,3,3)\wedge Q(3)\wedge 3>2 \wedge 3 \neq \nn)$. Since $\nn$ is treated as any other constant, we can compare it with $3$. By the {\em unique names assumption}, it holds $\nn \neq 3$.
\boxtheorem
\end{example}
Although our framework provides a precise semantics for conjunctive queries in $\nit{Conj}(\Sigma)$ or
$\nit{Conj}(\Sigma^\nulo)$, in both cases possibly containing (in)equalities involving $\nn$, a usual conjunctive query
in SQL should be first translated into a conjunctive query $\mc{Q}$ in  $\nit{Conj}^\sql(\Sigma^\nulo)$ if we want to
retain its intended semantics. After that $\mc{Q}^\nit{rw}$ can be computed.

\vspace{-2mm}
\section{Secrecy Instances}\label{sec:secInstance}
In this work  we will make use of {\em null} to protect secret information.  The basic idea that we develop in this and the
next sections is that the extensions of the secrecy views, obtained as query answers, should contain only the tuple with
{\em nulls}  or become empty. In this case we will say that {\em the view is null}.

\begin{definition} \label{def:null} A query $\mc{Q}(\bar{x})$ {\em is null} on instance $D$ if $\mc{Q}^N\!(D) \subseteq \{\langle \nn,\ldots,\nn\rangle \}$ (with the tuple inside with the same length as $\bar{x}$). A view $V(\bar{x})$ is
null on $D$ if the query defining it is null on $D$. \boxtheorem
\end{definition}

\vspace*{-9mm}
\begin{example}\label{ex:secView} (example \ref{ex:ansquery} continued)~Consider the secrecy view $\Vs(x)\leftarrow
R(x,y,z),$ $S(y),~y > 2$. Its corresponding FO query $\mc{Q}^{\Vs}(x)$ in the one in (\ref{eq:quer}), namely:

\vspace{1mm}\centerline{$\mc{Q}_2(x): \ \exists y \exists z(R(x,y,z)\wedge S(y) \wedge y>2).$}

\vspace{1mm}\noindent
Under the semantics of secrecy in the presence of {\em null}, we expect the view to be null.  This requires the  values
 for attribute $A$ associated with variable $x$ in $\mc{Q}_2$ to be $\nn$, or the values in $B$ associated with
 variable $y$ in $\mc{Q}_2$ to be $\nn$, or the negation of the comparison to be $\nit{true}$. These three cases
 correspond to the three assignments of Example \ref{ex:ansquery}.
 Thus, the view extension is $V_s(D_2) = \{\langle \nn \rangle \}$, which shows that the view is null on $D_2$.
 \boxtheorem
\end{example}
In this example we are in an ideal situation, in the sense that we did not have to change the instance to obtain a ``secret answer".  However, this may be an exceptional situation, and we will have to virtually ``distort" the given instance by replacing -as few as possible- non-null attribute values by $\nn$. More generally,
since it does not  necessarily holds that each secrecy becomes null on an instance $D$ at hand, the view extensions will be obtained from an alternative,
possibly virtual, version $D'$ of $D$ that does make each of those views null. In this sense, $D'$ will be an {\em admissible} instance (cf. Definition \ref{df:Newadmissible} below).  At the same time, we want $D'$ to stay as close as possible to $D$ (cf. Definition \ref{df:secrecyinstance} below).
Since there may be more that one such instance $D'$, we query all of them simultaneously, and return the
{\em certain answers} \cite{IT84} (cf. Definition \ref{def:secAns} below). Each of the query and view evaluations is
done according to the notion of $N$-answer introduced in Section \ref{app:nans}.

First,  we define the instances that make the secrecy views empty or null.

\begin{definition}\label{df:Newadmissible}
 An instance $D$ for schema $\Sigma$ is {\em admissible} for a set $\mc{V}^s$ of secrecy views of the form (\ref{eq:view})
if under the $N$-answer semantics (cf. Definition \ref{def:nvs}),  each $V_s(D)$ is empty or in all its tuples only {\em null} appears. $\nit{Admiss}(\mc{V}^s)$
 denotes the set of admissible instances. \boxtheorem
\end{definition}
As Example \ref{ex:secView}
shows, $D_2$ is admissible for the the given view. It also shows that there are some attributes that are particularly relevant for the view to be null, $A$ and $B$ in that case. In the following, we make precise this notion of {\em secrecy-relevant attribute}
(cf. Definition \ref{df:comAtts}(d) below). Before we used (plain) ``relevance" associated to variables for query answering
under nulls. Not surprisingly, the new notion is based on the previous one. This will allow us to provide an
alternative and more operational  characterization of secrecy instances (cf. Proposition \ref{prop:admis} below).

\begin{definition}  \label{df:comAtts} Consider a view $\Vs$ defined as in (\ref{eq:view}).\\  (a) For  $R \in \mc{R}$ in the body of (\ref{eq:view}) and a term $t$ (i.e. a variable or constant), $\nit{pos}^R\!(\Vs,t)$ denotes the set of {\em positions} in
$R$ where $t$ appears in the body of $\Vs$'s definition. \\
(b) The set of \emph{combination attributes} for $\Vs$ is:

\centerline{${\cal C}(\Vs)=$
$\{R[i]~|$ for a relevant variable $v$,  $i\in
\nit{pos}^R\!(\Vs,v)\}$.}

\noindent (c) The set of \emph{secrecy attributes} for $\Vs$  is: \ ${\cal S}(\Vs)=$ $\{R[i]~|$ for an $x$ in $V_s(\bar{x})$ in (\ref{eq:view}), $i
\in \nit{pos}^R\!(\Vs,v)\}$.\\
(d) The set of \emph{s-relevant attributes}\footnote{For distinction from the notion
 of relevant attribute/variable used in Sections \ref{sec:nvs} and \ref{app:nans}.} for a secrecy view $\Vs$ are those
(associated to positions) in
the set
${\cal A}(\Vs)$ $=$ ${\cal C}(\Vs)\cup {\cal S}(\Vs)$.
\boxtheorem
\end{definition}
Combination attributes for a secrecy view $\Vs$ are those involved in joins or built-in predicates (other than
built-ins with explicit $\nn$).
Secrecy attributes are those appearing in the head of $\Vs$'s definition, and accordingly, collect the query answers,
which are expected to be secret. Hence, ``secrecy attributes". They correspond to the free variables
in the associated query $\mc{Q}^{\Vs}$.

\ignore{
\begin{definition}\label{df: relAtts}  The set of \emph{s-relevant attributes}\footnote{For distinction from the notion
 of relevant attribute/variable used in Sections \ref{sec:nvs} and \ref{app:nans}.} for a secrecy view $\Vs$ are those
(associated to positions) in
the set
${\cal A}(\Vs)={\cal C}(\Vs)\cup {\cal S}(\Vs)$. \boxtheorem
\end{definition}
}

\begin{example}\label{ex:relevant} (example \ref{ex:secView} continued) Consider again the secrecy view $\Vs(x)\leftarrow
R(x,y,z),$ $S(y),~y > 2$. Here
  ${\cal C}(\Vs)=\{R[2],S[1]\}$, because $y$ is the only relevant variable; and  ${\cal S}(\Vs)=\{R[1]\}$, because
$x$ is the only free variable. In consequence,  ${\cal A}(\Vs)=\{R[1],$ $ S[1], R[2]\}$. Attribute $C$, i.e.
$R[3]$,
is not s-relevant. Actually, its value  is not  relevant to obtain the view extension.\boxtheorem
\end{example}

The following proposition provides a characterization of admissible instance for a set of secrecy of views in terms
of classic FO satisfaction (cf. \cite[Proposition 1]{li}). In it we use the notation $D \models \gamma$ for the
classic notion of satisfaction by an instance $D$ of FO formula $\gamma$, where {\em null} is treated as any other constant.

\begin{proposition} \label{prop:admis}
 Let $\mc{V}^s$ be a set of secrecy views, each of whose elements  $\Vs$ is of the form (\ref{eq:view}), and has an expression $\mc{Q}^{\Vs}\!(\bar{x})\!: \exists \bar y(\bigwedge_{i=1}^{n} R_i(\bar x_i)\wedge\varphi)$  as a conjunctive query. For an instance $D$, $D \in \nit{Admiss}(\mc{V}^s)$
iff for each $\Vs \in \mc{V}^s$, $D\models \mbox{\nit{Null}-}V^{\!s}$, where $\mbox{\nit{Null}-}V^{\!s}$ is the following  sentence associated to $\mc{Q}^{\Vs}$:
\begin{eqnarray}
\overline{\forall}(\bigwedge_{i=1}^{n} R_i(\bar
{x}_i) &\longrightarrow& \bigvee_{v \ \in \ \bigcup_i^n \! \! \bar{x_i} \ \cap \ {\cal C}(\Vs)}\!\!v = \nn~~\vee \label{eq:class}\\&& \bigwedge_{u \ \in \ \bigcup_i^n \! \! \bar{x_i} \ \cap \ {\cal S}(\Vs)}\!\!u = \nn~~\vee~ \neg \varphi ). \  \blacksquare \nonumber
\end{eqnarray}
\end{proposition}
\noindent In the theorem, $\overline{\forall}$ denotes the universal closure of the formula that follows it; and
$v \in (\bigcup_i^n \! \! \bar{x_i} \cap {\cal C}(\Vs))$ indicates that variable $v$ appears in some of the atoms
$R_i(\bar{x}_i)$ and in a combination
attribute, etc.

Sentence $\mbox{\nit{Null}-}V^{\!s}$ in (\ref{eq:class}) originates in the FO rewriting $(\mc{Q}^{\Vs})^\nit{rw}$ as in (\ref{eq:rew}) of the query $\mc{Q}^{\Vs}$ associated to $V^{\!s}$, and the requirement that the latter
becomes null on $D$.

\begin{example}~(example \ref{ex:relevant} continued)~ According to the above definition, in order to check whether the database instance $D_2$ is admissible, the following must hold:
\begin{eqnarray*}
D_2 \models \forall x \forall y\forall z (R(x,y,z)\wedge
S(y) \ &\longrightarrow& x = \nn \ \vee \\&&\hspace*{-7mm}y = \nn \ \vee \ y \leq 2).
\end{eqnarray*}
 When checking sentence on $D_2$, $\nn$  is treated as any other constant. Notice that the values for the
 non-s-relevant attributes do not matter.

 For $x=1, y=1$, the antecedent of the implication is satisfied. For these values, the
consequent is also
satisfied, because $y = 1 <2$.  For $x=2, y=\nn$, the consequent is satisfied since $y$ is $\nn$. For $x=\nn, \ y=3$, the antecedent  is satisfied. For these values, the consequent is also
satisfied, because $\nn = \nn$ is true. So, $D_2\models_N \mc{Q}^{\Vs}$, and instance $D_2$ is admissible. \boxtheorem
\end{example}
The next step consists in selecting from the admissible instances those that are close to the database we are protecting.
This requires introducing a notion of distance or an order relationship between instances for a same schema. This would allow us to talk about minimality of change.
Since, in order to enforce privacy on an instance $D$, we will virtually change attribute values by $\nn$, the comparison of instances has to take this kind of changes and the presence of  $\nn$ in tuples into account. Intuitively,
a \nit{secrecy instance} for $D$ will be admissible and  also minimally differ from $D$.

\begin{definition}
(a) The binary relation $\sqsubset$ on the database domain $\mc{U}$, is defined  as follows: \ $c \sqsubset d$ iff
$c = \nn$ and $d \neq \nn$. Its reflexive closure is  $\sqsubseteq$.\\
(b)
For $\bar{t}_1$ $=$ $\langle c_1,\ldots,c_n\rangle$ and $\bar{t}_2$ $=$ $\langle d_1,\ldots,d_n\rangle$
in $\mc{U}^n$: $\bar{t}_1\sqsubseteq \bar{t}_2$ iff $c_i \sqsubseteq d_i$ for each $i \in \{1,\ldots,n\}$. Also,  $\bar{t}_1 \sqsubset \bar{t}_2 $ iff $\bar{t}_1\sqsubseteq \bar{t}_2$ and $\bar{t}_1 \neq \bar{t}_2$.
 \boxtheorem
\end{definition}
This partial order relationship $\bar{t}_1 \sqsubseteq \bar{t}_2$ indicates that  $\bar{t}_1$ is less or equally informative than $\bar{t}_2$.
For example, tuple $(a,null)$ provides less information than tuple $(a,b)$. Then, $(a,\nn)$ $\sqsubset$ $(a,b)$ holds.
\ignore{$(a,\nn) \sqsubsetneqq (a,b)$ holds.}

In order to capture the fact that we are just modifying attribute values, but not inserting or deleting
tuples, we  will assume (sometimes implicitly) that database tuples have {\em tuple identifiers}. More precisely,  each predicate has an additional, first, attribute $\nit{ID}$, which is a
key for the relation, and whose values
are taken in $\mathbb{N}$ and not subject to changes. In consequence, tuples in an instance $D$ will be of the form $R(k,\bar{t})$, with $k \in \mathbb{N}$, and
$\bar{t} \in \mc{U}^n$, and $R \in \mc{R}$ is, implicitly,  of arity $n+1$.   Below, we will consider only instances $D'$ that are {\em correlated}
to $D$, i.e. there is a surjective function $\kappa$ from $D$ to $D'$, such that $\kappa(R(k,\bar{t})) =
R(k,\bar{t}')$, for some $\bar{t}'$. This mapping respects the predicate name and the tuple identifier. We say that $D'$ is $D$-correlated
(via $\kappa$). In the rest of this section, $D$ is a fixed instance, the one under privacy protection. We will usually
omit tuple identifiers.

\begin{definition} (a) For database tuples $R_1(k_1,\bar{t}_1),$ $R_2(k_2,\bar{t}_2)$: \
 $R_1(k_1,\bar{t}_1)$ $\sqsubseteq$ $R_2(k_2,\bar{t}_2)$ iff  $R_1 = R_2$, $k_1 = k_2$, and $t_1\sqsubseteq t_2$.\\
 (b) For instances $D_1, D_2$: \ $D_1 \sqsubseteq D_2$ iff for every tuple $R_1(k_1,\bar{t}_1)\in D_1$, there is
 a tuple $R_2(k_2,\bar{t}_2)$ with $R_2(k,\bar{t}_2)$  $\sqsubseteq$  $R_1(k,\bar{t}_1)$.\\
 (c) For $D$-correlated instances $D_1,$ $D_2$: \ $D_1$ $\leq_D$ $D_2$ iff: i. $D_1, D_2 \sqsubseteq D$, and ii. $D_2 \sqsubseteq D_1$.
\ignore{$R(k,\bar{t}_1)$ $\in D_1$,
 $R(k,\bar{t}_2)$ $\in D_2$ its holds $R(k,\bar{t}_2)$  $\sqsubseteq$  $R(k,\bar{t}_1)$.} As usual,
 $D_1$ $<_D$ $D_2$ iff $D_1$ $\leq_D$ $D_2$, but not $D_2$ $\leq_D$ $D_1$.
 \boxtheorem
\end{definition}
Notice that the condition (c)i. for the partial order $\leq_D$
forces $D_1$ and $D_2$ to be obtained from $D$ by updating attribute values by $\nn$. Condition (c)ii. inverts the partial order $\sqsubseteq$ between
tuples (and between instances).
The reason
 is that we want secrecy instances to be {\em minimal} wrt the {\em set of changes}  of attributes values by nulls (as customary for database repairs \cite{bertossi11}). Informally, when $D_1 \leq_D D_2$, $D_1$ is obtained from $D$, in comparison
with $D_2$, via ``less" replacements of values by nulls, and then is close to $D$.

\begin{definition} \label{df:secrecyinstance} An instance $D_{\!s}$ is a \emph{secrecy instance} for $D$ wrt a set  ${\cal V}^s$ of secrecy views iff: (a) $D_{\!s}$ $\in$ $\nit{Admiss}({\cal V}^s)$, and (b) $D_{\!s}$ is $\leq_D$-minimal in the class of $D$-correlated database instances that satisfy (a). (I.e. there is no instance $D'$ in that class with $D' <_D D_{\!s}$.) $\nit{Sec}(D,{\cal V}^s)$ denotes the set of all the secrecy instances for $D$ wrt $\mc{V}^s$.
\boxtheorem
\end{definition}
Notice that a secrecy instance nullifies all the secrecy views, is obtained from $D$ by changing attribute values by
$\nn$, and the set of changes is minimal wrt set inclusion.\footnote{As opposed to minimizing the cardinality of that set.
Cf. \cite{bertossi11} for a discussion of different forms of ``repairs" of databases.}

\begin{example} \label{ex:minimal} Consider the instance $D=\{P(1,2),R(2,1)\}$ for schema $\mc{R} =
\{P(A,B),$ $R(B,C)\}$. With tuple identifiers (underlined), it takes the form $D=\{P(\underline{1},1,2),$ $R(\underline{1},2,1)\}$.
Consider also the \emph{secrecy view}:

\vspace{1mm}
\centerline{$\Vs(x,z)\leftarrow P(x,y), R(y,z), \ y<3.\footnote{It would be easy to consider tuple ids
in queries and view definition, but they do not contribute to the final result and will only complicate the notation. So, we
skip tuple ids whenever possible.}$}

\vspace{1mm}\noindent
$D$ itself is not admissible (it does not nullify the secrecy view), and then it is not a secrecy instance either. Now,
consider the following alternative updated instances $D_i$:

{\small \begin{center}
 \begin{tabular}{|c|l|}
  \hline
  $D_1$ & $\{P(\underline{1},\nn,2),R(\underline{1},2,\nn)\}$\\
  $D_2$ & $\{P(\underline{1},1,\nn),R(\underline{1},2,1)\}$ \\
  $D_3$ & $\{P(\underline{1},1,2),R(\underline{1},\nn,1)\}$ \\
  $D_4$ & $\{P(\underline{1},1,\nn),R(\underline{1},\nn,1)\}$ \\
  \cline{1-2}
 \end{tabular}
\end{center} }

\noindent For example, for $D_1$ the set of changes can be identified with the set of changed positions: \
$U_1 = \{P[1], R[2]\}$ ($\nit{ID}$ has position $0$). The $D_i$ are all admissible,  that is (cf. (\ref{eq:class})):
\begin{eqnarray*}
D_i &\models&  \forall x \forall y \forall z (P(x,y)\wedge R(y,z) \ \longrightarrow\\&& \!\!\!(y = \nn \vee (x = \nn \wedge z = \nn) \vee y \geq 3).
\end{eqnarray*}
$D_1$, $D_2$, and $D_3$ are the only three secrecy instances, i.e. they are $\leq_D$-minimal: The sets of changes
 $U_1$, $U_2 = \{P[2]\}$, and $U_3 = \{R[1]\}$ are all incomparable under set inclusion.
 $D_4$ is not minimal, because $U_4 = \{P[2],R[1]\} \supsetneqq U_3$, which is also reflected in the fact that
 $P(\underline{1},1, \nn)$ $\sqsubset P(\underline{1},1, 2)$; and then, $D_3 <_D D_4$. \boxtheorem
\end{example}

\vspace{-3mm}
\section{Privacy Preserving Query Answers}\label{sec:secretAns}
 Now we want to define and compute the {\em secret answers to queries} from a given database $D$ that is subject to privacy
 constraints, as represented by the nullification of the secrecy views. They will be defined on the basis of the class of
 secrecy instances for $D$. This class will be queried  instead of directly querying $D$.  In  this sense, we may consider
the class of secrecy instances as representing a {\em logical database}, given through its models. In such a case, the
intended answers are those that are true of all the  instances in the class, and become the so-called {\em certain answers}
\cite{IT84}.
\begin{definition}  \label{def:secAns} Let $\mc{Q}(\bar{x}) \in \nit{Conj}(\Sigma^\nulo)$. A tuple $\bar{c}$ of
constants in $\mc{U}$ is a {\em secret answer} to $\mc{Q}$ from
$D$ wrt to a set of secrecy views ${\cal V}^s$ iff $\bar{c} \in \mc{Q}^{N\!}(\ds)$ for each  $\ds
\in \nit{Sec}(D,{\cal V}^s)$. ${\nit SA}(\mc{Q},D, \mc{V}^s)$ denotes the set of all secret answers. \boxtheorem
\end{definition}
\vspace{-5mm}\begin{example} (example \ref{ex:minimal} continued). Consider the query $\mc{Q}(x,z)\!:
\exists y(P(x,y)\wedge R(y,z)\wedge y<3)$. According to Definition \ref{def:nvs}, it holds:
 $\mc{Q}^{N\!}(D_1)=\{\langle \nn,\nn\rangle\}$, $\mc{Q}^{N\!}(D_2)=\emptyset$, and $\mc{Q}^{N\!}(D_3)=\emptyset$.
These answers can also be obtained by first rewriting
 $\mc{Q}$, as in (\ref{eq:rew}), into the query
 $\mc{Q}^\nit{rw\!}(x,z): \exists y(P(x,y)\wedge R(y,z) \wedge y<3 \wedge y\neq \nn)$, which can be
 evaluated on each of the secrecy instances treating
 \nit{null} as any other constant.

We obtain $\nit{SA}(\mc{Q},D, \{V_{\!s}\})= \mc{Q}^{N\!}(D_1) \cap \mc{Q}^{N\!}(D_2) \cap \mc{Q}^{N\!}(D_3) = \emptyset$.
 This is as expected, because in this example, $\mc{Q}$ is $\mc{Q}^{\Vs}$, the query associated to the secrecy view.
\boxtheorem \end{example}
The idea behind answering queries from the secrecy instances (SIs) for $D$ is that the answers are still close to those we would
have obtained from $D$ (because SIs are maximally close to $D$). Furthermore, since all
 the secrecy views become null on the SIs, the answers returned to any query, not necessarily to a secrecy view computation,
 will take this property into account. In the query answering part we are using a {\em skeptical or cautious
semantics}, that sanctions as true what is simultaneously true in a whole class of models, or instances in our case
(the SIs). Now we analyze to what extent this approach does protect the sensitive
data. A restricted  user may try to pose several queries to obtain sensitive information.

\begin{example}\label{ex:sd} Consider  instance $D=\{P(1,2),$ $P(3,4),$ $R(2,1),$ $R(3,3)\}$ for schema $\mc{R} =
\{P(A,B), R(B,C)\}$, and the secrecy view \ $\Vs(x,z)\leftarrow P(x,y),R(y,z)$. In this case, $\Vs^{N\!}(D) =
\{\langle 1,1\rangle\}$. \ $D$ has the following SIs:

{\small \begin{center}
 \begin{tabular}{|c|l|}
  \hline
  $D_1$ & $\{P(\nn,2),P(3,4),R(2,\nn),R(3,3)\}$\\
  $D_2$ & $\{P(1,\nn),P(3,4),R(2,1),R(3,3)\}$ \\
  $D_3$ & $\{P(1,2),P(3,4),R(\nn,1),R(3,3)\}$ \\
  \cline{1-2}
 \end{tabular}
\end{center}}

\noindent The user may pose the queries $\mc{Q}_1(x,y)\!: P(x,y)$ and $\mc{Q}_2(x,y)\!: R(x,y)$,
trying to reconstruct  $D$.  It holds  $\mc{Q}_1^{\!N\!}(D_1)$ $=$
$\{\langle \nn,2\rangle,$ $\langle 3,4\rangle\}$, $\mc{Q}_1^{\!N\!}(D_2)$ $=$ $\{\langle 1,\nn\rangle,$
$\langle 3,4\rangle\}$, $\mc{Q}_1^{\!N\!}(D_3)$ $=$ $\{\langle 1,2\rangle,$ $\langle 3,4\rangle\}$. Then,
$\nit{SA}(\mc{Q}_1,D,\{\Vs\})$ $=$ $\{\langle 3,4\rangle\}$.
Now,  $\mc{Q}_2^{\!N\!}(D_1) = \{\langle 2,\nn\rangle,\langle 3,3\rangle\}$, $\mc{Q}_2^{\!N\!}(D_2) = \{\langle 2,1\rangle,$
$\langle 3,3\rangle\}$,
$\mc{Q}_2^{\!N\!}(D_3) =$ $\{\langle \nn,1\rangle,$ $\langle 3,3\rangle\}$. Then, $\nit{SA}(\mc{Q}_2,$ $D,\{\Vs\}) =\{\langle 3,3\rangle\}$.

By combining the secret answers to  $\mc{Q}_1$
and $\mc{Q}_2$,  it is not possible to obtain $\Vs^{N\!}(D)$.
For the user who poses the queries $\mc{Q}_1$ and $\mc{Q}_2$, the
relations look as follows:
{\small \begin{center}
\begin{tabular}{c|c|c|} \hline $P$ & $A$ & $B$\\
                        \hline     & $3$ & $4$\\ \hhline{~--}
\end{tabular}
~~
\hspace{5mm}\begin{tabular}{c|c|c|} \hline $R$ & $B$ & $C$\\
                        \hline     & $3$ & $3$\\ \hhline{~--}
\end{tabular}
\end{center}}
\ignore{\noindent In this case, any other conjunctive query posed to detect the presence of initial nulls,  like
$\exists y (P(x,y)\wedge x=\nn)$, will get an empty set of secret answers, and the user will not know anything more about the contents
of the original instance.}  \vspace{-5mm}\boxtheorem
\end{example}
Now, we establish in general the impossibility of obtaining the contents of the secrecy views through the use of
secret answers to atomic queries (as in the previous example). Open atomic queries are the ``broader" queries we may
ask;  other queries are obtained from them by conjunctive combinations.

\begin{definition}\label{sec:sai}  Let $\cVs$ be a set of secrecy views $\Vs$. The {\em secrecy answer instance} for $\cVs$ from $D$ is
 $D_{\mc{V}^{\!s}} = \{R(\bar{c})~|~ R\in  \mc{R}$ and $\bar{c} \in \nit{SA}(R(\bar{x}),D, \cVs)\}$.\boxtheorem
\end{definition}
Here, we are building a database instance by collecting the secret answers (SAs)
to all the atomic queries of the form $\mc{Q}(\bar{x})\!: R(\bar{x})$, with $R \in {\cal R}$. This instance has the
same schema as  $D$.

\begin{example} (example \ref{ex:sd} continued)
Consider the secrecy view $\Vs(x,z) \leftarrow P(x,y),$ $R(y,z)$. It holds:
$D_{\{\Vs\}} = \{P(3,4)\}$ $\cup$ $\{R(3,3)\} = \{P(3,4),R(3,3)\}$. Notice that $\Vs^{\!N\!}(D_{\{\Vs\}})
= \emptyset = \nit{SA}(\mc{Q}^{\Vs},D,\{\Vs\}) = \bigcap_{i=1}^3 (\mc{Q}^{\Vs})^{\!N\!}(D_i) = \{\langle \nn, \nn \rangle\}
\cap \emptyset \cap \emptyset$.
\boxtheorem
\end{example}
\begin{proposition}\label{prop:leakage}  For every $\Vs$ of the form (\ref{eq:view}) in $\cVs$, $SA(\mc{Q}^{\Vs},D,\cVs)= \Vs(D_{\mc{V}^{\!s}})$.
\boxtheorem
\end{proposition}
This proposition tells us that by combining SAs to queries, trying to reconstruct the original instance, we cannot obtain
more information that the one provided by the SAs (cf. \cite[Proposition 2]{li} for a proof).

The original database $D$ may contain null values, and users have to count on that.  A restricted user will receive as query
answers the SAs, which are defined and computed through null values. This user could obtain
nulls from a query, and hopefully he will not know if they were already in $D$ or were (virtually) introduced for
privacy purposes. This is fine and accomplishes our goals. However,  as long as the user does not have other kind of
information.

\begin{example} Consider the instance $D=\{P(1,1)\}$, and the secrecy view $\Vs(x)\leftarrow P(x,y),x=1$.  $D$ has
only one secrecy instance $D_s$:

{\small \begin{center}
\begin{tabular}{c|c|c|}
  \hline
  $P$ & $A$ & $B$ \\ \hline
   & $\nn$ & $1$ \\
  \cline{2-3}
 \end{tabular}
\end{center}}

\noindent For the query $\mc{Q}(x): \exists y(P(x,y)\wedge x=1)$ associated to the secrecy view, the secrecy answer to
$\mc{Q}(x)$ on $D$ is $\emptyset$. Now, the secrecy answer to $\mc{Q}'(x): \exists y P(x,y)$ is $\{\langle \nn \rangle\}$.
A user who receives this answer will not know if the null value was introduced to protect data.

However, if the user knows from somewhere else that there is an SQL's {\tt NOT NULL} constraint or a key constraint
on the first attribute, and that it is satisfied by $D$, then he will know that the received null was not originally
in $D$.
Furthermore, that it is replacing a non-null value. If he also knows that there is exactly one tuple in the
relation (a {\tt COUNT} query), and also the secrecy view definition,  he will infer that
$\langle 1 \rangle \in \Vs^{\!N\!}(D)$.
 \boxtheorem
\end{example}
In summary, for our approach to work, we rely on the following assumptions:
\begin{itemize}
\item [(a)] The user interacts via conjunctive query answering with a possibly incomplete database, meaning that the latter
may contain null values,
and this is something the former is aware of, and can count on (as with databases used in common practice). In this way,
if a query returns answers with null values, the user will not know if they were originally in the database or were
introduced for protection at query answering time.
\item [(b)] The queries request data, as opposed to schema elements,
like integrity constraints and view definitions. Knowing the ICs (and about their satisfaction) in combination
    with query answers could easily expose the data protection policy. The most clear example is the one of a {\tt NOT NULL} SQL constraint, when we see
    nulls where there should not be any.
\item [(c)] In particular, the user does not know the secrecy view definitions. Knowing them would basically reveal
the data that is being protected and how.
\end{itemize}
These assumptions are realistic and make sense in many scenarios, for example, when the database is being accessed through the web,
 without direct interaction with the DBMS via complex SQL queries, or through an ontology that offers a limited
 interaction layer. After all, protecting data may require additional measures, like withholding from certain users
 certain information that is, most likely, not crucial for many applications. From these assumptions and  Proposition
 \ref{prop:leakage}, we can conclude that the user cannot obtain information about the secrecy views through a
 combination of SAs to  conjunctive queries. Therefore, there is not leakage of sensitive information.

\vspace{-3mm}
\section{Secrecy Instances and Logic Programs}\label{sec:logicProm}
The updates leading to the secrecy instances (SIs) should not physically change the database. Also, different users may be
restricted by different secrecy views. Rather, the possibly several SIs have to be virtual, and used mainly as an auxiliary
notion  for the secret answer semantics.  We expect be able to avoid computing all the SIs, materializing them, and then
cautiously querying the class they form. We would rather
stick to the original instance, and use it as it is to obtain the secret answers.

One way to approach this problem
is via query rewriting. Ideally, a query $\mc{Q}$ posed to $D$ and expecting secret answers should be rewritten
into another query $\mc{Q}'$. This new query would be posed to $D$, and the usual answers returned by $D$ to $\mc{Q}'$
should be the secret answers to $\mc{Q}$. We would like $\mc{Q}'$ to be still a simple query, that can be easily
evaluated. For example, if $\mc{Q}'$ is FO, it can be evaluated in polynomial time in data. However, this possibility
is restricted by the intrinsic complexity
of the problem of computing or deciding secret answers, which is likely to be higher than polynomial time in data
(cf. Section \ref{sec:cqa}). In consequence, $\mc{Q}'$ may not even a FO query, let alone conjunctive.

An alternative approach is to specify the SIs in a compact manner, by means of a logical theory, and do
reasoning from that theory, which is in line with skeptical query answering. This will not decrease a
possibly high
intrinsic complexity, but can be much more efficient
than  computing all the secrecy instances and querying them in turns.
Wrt the kind of logical specification needed, we can see that secret query answering (SQA) is a {\em non-monotonic} process.

\begin{example} Consider
$D= \{P(a)\}$, the secrecy view $V(x) \leftarrow P(x), R(x)$, and the query
$\mathcal{Q}: \ \nit{Ans}(x) \leftarrow P(x)$.
Here, $V(D) = \emptyset$, and then, $D$ itself is its only SI. Therefore,
$\nit{SA}(\mathcal{Q},D,\{V\}) = \{\langle a\rangle\}$.

Let us update $D$ to $D_1 = \{P(a), R(a)\}$. Now, $V(D_1)$ $=$ $\{\langle a\rangle\}$. The SIs for $D_1$ are:
$D_1'=\{P(\nit{null}), R(a)\}$ and $D_1''= \{P(a), R(\nit{null})\}$. It holds,
$\mathcal{Q}(D_1') = \{\langle \nn\rangle\}$ and $\mathcal{Q}(D_1'') = \{\langle a\rangle\}$. Then,
$\nit{SA}(\mathcal{Q},D_1,\{V\}) = \emptyset$. The previous secret answer is lost.\boxtheorem
\end{example}
The non-monotonicity of SQA requires a non-monotonic formalism to logically specify the
SIs of a given instance.
Actually, they can be specified as the stable models of a
disjunctive logic program, a so-called {\em secrecy program}.

Secrecy programs use annotation constants with the intended, informal semantics
shown in the table below. More precisely, for each database predicate $R \in
\mc{R}$, we introduce a copy of it with  an extra, final attribute (or argument) that contains
an annotation constant. So, a tuple of the form $R(\bar{t})$ would become an annotated atom of the
form $R(\bar{t},\mathbf{a})$.\footnote{We should use a new predicate, e.g. $R'$, but to keep the notation
simple, we will reuse the predicate. We also omit tuple ids.}  The  annotation constants are
used to keep track of virtual updates, i.e. of old and new tuples:

{\small \begin{center}
\begin{tabular}{|l|l|l|}
 \hline
  Annotation & Atom & The tuple $R(\bar a)$ ... \\
  \hline
 \hspace*{4mm}  $\au$ &  $R(\bar{a}', \au)$ & is  being updated  \\
\hspace*{4mm}  $\bu$ &  $R(\bar{a}, \bu)$ &  has been updated \\
\hspace*{4mm}  $\bft$ &  $R(\bar{a}, \bft)$ & is new or old  \\
 \hspace*{4mm} $\s$ & $R(\bar{a}, \s)$ & stays in the secrecy instance\\
  \hline
\end{tabular}
\end{center} }

\noindent In $R(\bar{a}, \bu)$, annotation $\bu$ means that the atom $R(\bar{a})$ has already been updated, and $\au$
should appear in the new, updated atom, say $R(\bar{a}',\au)$.
For example, consider a tuple $R(a,b)\in D$. A new tuple $R(a,\nn)$ is obtained by updating $b$ into $\nn$.
Therefore, $R(a,b,\bu)$ denotes the old atom before updating, while $P(a,\nn,\au)$ denotes the new atom after the update.

The logic program uses these annotations to go through different steps, until its stable models are computed. Finally,
the atoms needed to build an SI are read off by restricting a model of the program to atoms with the annotation $\s$.
As expected, the official semantics of the annotations is captured through the logic program; the table above is just for
motivation. In Section \ref{sec:app} we  provide the general form of $\Pi(D,\mc{V}^s)$,  the {\em secrecy logic program}
that specifies the SIs for an instance $D$ subject to set of secrecy views $\mc{V}^s$. The following example illustrates
the main ideas and issues.

\begin{example} (example \ref{ex:minimal} continued) \label{ex:logicP} Consider  $\mc{R}$ $=$
$\{P(A,B),$ $R(B,C)\}$, $D=$ $\{P(1,2),$ $R(2,1)\}$ and the secrecy
view \ $\Vs(x, z)\leftarrow P(x,y),R(y,z),y<3$.

The secrecy instance program $\Pi(D, \{V_s\})$ is as follows:

\noindent 1.~ $P(1,2)$. \ \ $R(2, 1).$ \ \ \ \  (initial database)

\vspace{1mm}
\noindent 2.~ $P(\nn,y,\au) \vee P(x,\nn,\au) \vee R(\nn,z,\au)$

\hfill $\leftarrow$ $P(x,y,\bft),~ R(y,z,\bft),y<3,~ y \neq \nn, \nit{aux}(x,z).$

\vspace{1mm}
~$R(y,\nn,\au)  \vee P(x,\nn,\au) \vee R(\nn,z,\au)$

\hfill $\leftarrow$ $P(x,y,\bft),~ R(y,z,\bft),y<3,~ y \neq \nn, \nit{aux}(x,z).$

\vspace{1mm}
~ $\nit{aux}(x,z)\leftarrow P(x,y,\bft), ~R(y,z,\bft),y<3,x \neq \nn.$\\
\\
~ $\nit{aux}(x,z)\leftarrow P(x,y,\bft), ~R(y,z,\bft),y<3,z \neq \nn.$

\vspace{-4mm}
\begin{eqnarray*}
3.~ P(x,y,\bu)\!\!\!\! &\leftarrow& \!\!\!\!P(x,y,\bft), R(y,z,\bft),y<3, y\neq \nn,\\
&&\nit{aux}(x,z),P(\nn,y,\au), x \neq \nn.\\
R(y,z,\bu)\!\!\!\! &\leftarrow& \!\!\!\!P(x,y,\bft), R(y,z,\bft),y<3, y\neq \nn,\\
&&\nit{aux}(x,z),R(y,\nn,\au), z\neq \nn.\\
P(x,y,\bu)\!\!\!\! &\leftarrow& \!\!\!\!P(x,y,\bft), R(y,z,\bft), y<3, y\neq \nn,\\
&&\nit{aux}(x,z),P(x,\nn,\au).\\
R(y,z,\bu)\!\!\!\! &\leftarrow& \!\!\!\!P(x,y,\bft), R(y,z,\bft), y<3, y\neq \nn,\\
&&\nit{aux}(x,z),R(\nn,z,\au).\\
4.~ P(x,y,\bft) &\leftarrow& \!\!\!P(x,y). \hspace{8mm} P(x,y,\bft)\leftarrow P(x,y,\au).\\
R(x,y,\bft)&\leftarrow& \!\!\!R(x,y). \hspace{8mm} R(x,y,\bft)\leftarrow R(x,y,\au).\\
5.~ P(x,y,\mathbf{s}) &\leftarrow& \!\!\!P(x,y,\bft),~\nit{not} ~P(x,y,\bu).\\
        R(x,y,\mathbf{s}) &\leftarrow& \!\!\!R(x,y,\bft),~\nit{not} ~R(x,y,\bu).
        \end{eqnarray*}
The facts  in 1. belong to the initial instance $D$, and become annotated right away with $\bft$ by rules 4. The most
important rules of the program are those in  2. and 3. They enforce the update semantics of secrecy in the presence of
$\nit{null}$ and
using $\nit{null}$. Rules in 2. capture in the body the violation of secrecy (i.e. a non-null view contents); and in the head, the intended way of
restoring secrecy: We can either update a combination of (combination) attributes or single
secrecy attributes with $\nn$. In this example, we need to update, with $\nn$, values in attribute $B$  or
in attributes $A$ and $C$, simultaneously.

 Since disjunctive programs do not allow conjunctions in the head, the intended head
 $(P(\nn,z) \wedge P(y,\nn)) \vee P(x,\nn)\vee Q(\nn,z) \leftarrow \nit{Body}$ is represented by means of two rules, as in 2.:
 $P(\nn,z) \vee P(x,\nn)\vee Q(\nn,z) \leftarrow \nit{Body}$ and
 $P(y,\nn) \vee P(x,\nn)\vee Q(\nn,z) \leftarrow \nit{Body}$.

Furthermore, we need to restore secrecy only if the given database is not already a secrecy instance, which happens
when the combination attribute $B$ is not null, the secrecy attributes $A$ and $C$ are not null, and formula
$\varphi$ is true. Predicate  $\nit{aux}(x,z)$ defined in 2. captures the condition  $\nit{not}~(x\neq\nn \wedge z\neq\nn)$.

The rules in 3. collect the tuples in the database that have already been updated and (virtually) no longer  exist in the
database. Rules 4. annotate the original the atoms and also the new version of updated atoms. Rules in 5. collect the tuples
that stay in the final state of the updated database: They are original or new, but have never been updated.\boxtheorem
\end{example}
The secrecy instances are in one-to-one correspondence with the restrictions to $\mathbf{s}$-annotated atoms of the stable
models of $\Pi(D, \mc{V}^s)$.\footnote{The proof of this claim is rather long,
and is similar in spirit to the proof of the fact that database repairs wrt integrity constraints \cite{bertossi06}
can be specified
by means of disjunctive logic programs with stable model semantics (cf. \cite{bravo07,barcelo02}).}

\begin{example}\label{ex:logicP2} (example \ref{ex:logicP} continued) The program has three stable models (the facts in 1.
are omitted):

\vspace{2mm}
\noindent $M_1 =\{P(1,2,\bft),~ R(2,1,\bft),~ aux(1,1), \underline{P(1,2,\s)},$\\ \phantom{po}\hfill $R(2,1,\bu), R(\nn,1,\au),
R(\nn,1,\bft),\underline{R(\nn,1,\s)}\}$.

\noindent
$M_2= \{P(1,2,\bft),~ R(2,1,\bft), ~\nit{aux}(1,1), P(1,2,\bu),$\\ \phantom{po} \hfill $\underline{R(2,1,\s)},P(1,\nn,\au),
P(1,\nn,\bft), \underline{P(1,\nn,\s)}\}$.

\noindent
$M_3=\{P(1,2,\bft),~ R(2,1,\bft),~ \nit{aux}(1,1), P(1,2,\bu),$\\ \phantom{po} \hfill $R(2,1,\bu),P(\nn,2,\au),
R(2,\nn,\au),P(\nn,2,\bft),$\\ \phantom{po} \hfill $R(2,\nn,\bft), \nit{aux}(1,null),
\nit{aux}(null,1),\underline{P(\nn,2,\s)},$\\ \hspace*{6mm}$\underline{R(2,\nn,\s)}\}$.

\vspace{2mm}
\noindent The secrecy instances  are built by selecting the underlined atoms, obtaining: $D_1$ $=$
$\{P(1,2),$  $R(\nn,1)\}$,  $D_2$ $=$ $\{P(1,\nn),$ $R(2,1)\}$, and $D_3$ $=$
$\{P(\nn,2),$  $R(2,\nn)\}$. They coincide with those in Example \ref{ex:minimal}.
\boxtheorem
\end{example}
In order to compute secret answers to a query, it is not necessary to
explicitly compute all the stable models. Instead, the query can be posed directly on top of the program and
answered according to the skeptical semantics. This will return the secret answers to the query.
The query has to be formulated as a top-layer program, with $\s$-annotated atoms, that are those that affect the query.
A system like {\em DLV} can be used. It computes the disjunctive stable-model semantics, with an
interface to commercial DBMSs \cite{dlv}.
\begin{example}\label{ex:logicP2Q} (example \ref{ex:logicP2} continued)
We want the secret answers to the conjunctive query

\vspace{1mm}
\centerline{$\mc{Q}(x,z): \exists y (P(x,y)\wedge R(y,z)\wedge y<3).$}

\vspace{1mm}\noindent
This requires first rewriting it, as in (\ref{eq:rew}), into $\mc{Q}^\nit{rw}(x,y)\!: \exists y (P(x,y)\wedge R(y,z)\wedge y<3
\wedge y \neq \nn)$. This new query can be evaluated against instances with \nit{null} treated as any other constant.
In its turn, $\mc{Q}^\nit{rw}$ is transformed into a query program
with all the database atoms using annotation $\s$:

\vspace{1mm}
\centerline{$\nit{Ans}(x,z)\leftarrow P(x,y,\s),R(y,z,\s), \ y<3, \ y\neq\nn.$}

\vspace{1mm}\noindent This one is evaluated in combination with
the secrecy program
in Example \ref{ex:logicP}, under the skeptical semantics. In this evaluation, \nit{null} is treated as an ordinary constant.
\boxtheorem
\end{example}

\vspace{-3mm}
\subsection{The general secrecy logic program}\label{sec:app}
  To provide the general form  of secrecy logic program, we need to introduce some notation first. We recall that our view definitions are of the form
\begin{equation} \Vs(\bar x) \leftarrow R_1(\bar{x}_1), \ldots,
R_n(\bar{x}_n), \ \varphi. \label{eq:newView}
\end{equation}
Some of the variables\footnote{To be more precise, we should talk about variables in relevant
 positions or arguments, as we did before, e.g. in Section \ref{sec:secInstance}, but the description would be less
 intuitive.}in atoms in the body of the definitions are relevant, as in Definition \ref{df:comAtts}, and their values will be
 replaced by $\nn$. As expected, and illustrated in Example \ref{ex:minimal}, those atoms and variables
 play a crucial role in the program.

 For an atom of the form $R(\bar{x})$ and variables $\bar{y} \subseteq \bar{x}$, $R(\bar{x})\frac{\bar{y}}{\nn}$ denotes
  $R(\bar{x})$ with all the variables in $\bar{y}$ replaced by $\nn$. In reference to (\ref{eq:newView}), with this
  notation, we define:
\begin{eqnarray*}
  \mc{CP}(\Vs) \!\! &=& \! \! \{R_i(\bar{x}_i)\frac{\bar{y}}{\nn}~|~R_i(\bar{x}_i) \mbox{ is in body of (\ref{eq:newView})},\\ &&~~~\bar{y}=\{y_1,...,y_n\} \subseteq \bar{x}, \mbox{ and } y_i \in \mc{C}(\Vs))\}.\\
  \mc{SP}(\Vs) \!\!&=& \! \! \{R_i(\bar{x}_i)\frac{\bar{y}}{\nn}~|~R_i(\bar{x}_i) \mbox{ is in body of (\ref{eq:newView})},\\ &&~~~\bar{y}=\{y_1,...,y_n\} \subseteq \bar{x}, \mbox{ and } y_i \in \mc{S}(\Vs))\}.
  \end{eqnarray*}
\ignore{ for each $R_i(\bar x_i)$ in the body of (\ref{eq:newView},
  if $(\bar x_i \cap \mc{C}(\Vs)) \neq\emptyset$, let $\mc{CP}(\Vs)=\{\nit{change}(R_i(\bar x_i),y,\nn),$ for each $y\in (\bar x_i\cap \mc{C}(\Vs))\}$, and
  if
  $(\bar x_i\cap \mc{S}(\Vs))\neq\emptyset$, let $\mc{SP}(\Vs)=\{\nit{change}(R_i(\bar x_i),\bar y,\nn),$ for  $\bar y = (\bar x_i\cap \mc{C}(\Vs))\}$.
  }
For the sets of predicate positions,  $\mc{C}(\Vs)$ and $\mc{S}(\Vs)$, see Definition \ref{df:comAtts}.
The atom sets $\mc{CP}(\Vs)$ and $\mc{SP}(\Vs)$ will be used in the head of the disjunctive rules that
change some relevant attribute values into nulls (rules 2. in Example \ref{ex:minimal}).

\begin{example} For the secrecy view $\Vs(x,z,w)\leftarrow P(x,y),Q(y,z,w)$, it holds: $\mc{C}(\Vs) = \{P[2], Q[1]\}$ and
$\mc{S}(\Vs)$
$=$ $ \{P[1],Q[2],Q[3]\}$. Thus, $\mc{CP}(\Vs)=\{P(x,\nn), Q(\nn,z,w)\}$, and
$\mc{SP}(\Vs)=\{P(\nn,y),$ $Q(y,\nn,\nn)\}$. \boxtheorem
\end{example}
Given a database instance $D$, a set $\mc{V}^s$ of secrecy views $\Vs$s,
each of them of the form (\ref{eq:newView}),
the secrecy program $\Pi(D, \mc{V}^s)$ contains the following rules:

\vspace{2mm} \noindent
1. Facts: $R(\bar{c},\bft)$ for each atom $R(\bar{c}) \in D$.

\vspace{2mm} \noindent
2. For every $\Vs$ of the form (\ref{eq:newView}), if $\mc{SP}(\Vs) =$ $\{R^1(\bar{x}_1),$ $\ldots,R^a(\bar{x}_a)\}$, and $\mc{CP}(\Vs)$ $=$ $\{R^1(\bar{x}_1),$ $...,R^b(\bar{x}_b)\}$, then the program contains the rules:

\noindent (a) If $\mc{S}(\Vs)\cap \mc{C}(\Vs) \not = \emptyset$, the rule:

 \vspace{1mm}\noindent                   $\bigvee\limits_{R^{c} \in \mc{CP}(\Vs)} \hspace{-4mm}R^{c}(\bar{x}_{c},\au) \
 \leftarrow \ \bigwedge_{i=1}^n R_i(\bar{x}_i,\bft), \
                    \varphi, \!\!\! \bigwedge\limits_{v_l \in \mc{C}(\Vs)} \!\!\! v_l \not = \nn$.

 \vspace{1mm}\noindent (b) If $\mc{S}(\Vs)\cap \mc{C}(\Vs) = \emptyset$, for each $R^d\in \mc{SP}(\Vs)$, $1\leq d\leq a$,
 the rule:

 \vspace{1mm}\noindent    $R^d(\bar{x}_{d}, \au) \vee \hspace{-3mm} \bigvee\limits_{R^c \in \mc{CP}(\Vs)}
 \hspace{-3mm}R^c(\bar{x}_{c},\au) \ \leftarrow \ \bigwedge_{i=1}^n R_i(\bar{x}_i,\bft), \ \varphi,$

                   \hspace*{3.7cm} $\bigwedge\limits_{v_l \in \mc{C}(\Vs)} \vspace{-3mm} v_l \neq \nn,  \
                   \nit{aux}_{\Vs}(\bar{x}).$

\vspace{3mm} \noindent Plus rules defining the auxiliary predicates: If $\mc{S}(\Vs)=\{x^1,...,x^k\}$ and $\bar{x}=
\langle x^1, \ldots, x^k\rangle$,
then for each $1 \leqslant i \leqslant k$, the rule

\vspace{1mm}  \hspace*{6mm} $\nit{aux}_{\Vs}(\bar{x}) \leftarrow \bigwedge_{i=1}^n R_i(\bar{x}_i,\bft)\wedge \varphi \wedge x^i \neq
\nn.$

\vspace{2mm} \noindent
3. The old tuple collecting rules:

\noindent
(a) For each $R^j\in \mc{SP}(\Vs)$, $1\leq j\leq a$:

\vspace{1mm} $R^j(\bar{x}_j,\bu) \leftarrow \bigwedge_{i=1}^n R_i(\bar{x}_i,\bft), \ \varphi, \ \nit{aux}_{\Vs}(\bar{x}),$

        \hspace*{12mm}$\bigwedge\limits_{v_l \in \mc{C}(\Vs)} \hspace{-3mm}v_l \neq \nn, \ R^j(\bar{x}_{j},\au),
        \hspace{-3mm}  \bigwedge\limits_{v_l \in \mc{S}(\Vs)\cap \bar{x}_{j}} \hspace{-3mm}v_l \neq \nn.$

\vspace{1mm}\noindent
(b)  For each $R^c \in \mc{CP}(\Vs)$, $1\leq c\leq b$:

\vspace{1mm}   $R^c(\bar{x}_{c},\bu) \leftarrow \bigwedge_{i=1}^n R_i(\bar{x}_i,\bft), \ \varphi, \
\nit{aux}_{\Vs}(\bar{x}),$

\hspace*{3.5cm}$\bigwedge\limits_{v_l \in \mc{C}(\Vs)} \hspace{-3mm}v_l \neq \nn, \ R^c(\bar{x}_{c},\au).$

\vspace{2mm} \noindent
4. For each  $R \in \mc{R}$, the  rule: \ \
        $R(\bar{x},\bft)\leftarrow R(\bar{x},\au)$.

\vspace{2mm} \noindent
5. For each  $R \in \mc{R}$, the rule:

     \hspace*{2cm}   $R(\bar{x},\s)\leftarrow R(\bar{x},\bft),\n R(\bar{x},\bu)$.

\vspace{2mm}
\noindent Rules in 1. create program facts from the initial instance. Rules in 2. are the most important and
express how to impose secrecy by changing attribute values into nulls. Notice that, by definition,
$\mc{CP}(\Vs)$ and $\mc{SP}(\Vs)$
already already include those changes. The body of the rule becomes true when the database instance does not nullify the
view, and the head captures the intended ways of imposing secrecy. Rules in 3. collect the tuples in the database that have already been updated and (virtually) no longer exist in the database. Rules 4. capture the atoms that are part
of the database or updated atoms in the process of imposing secrecy. Rules in 5. collect the tuples in
the secrecy instance, as those that did not become old.

The same secrecy program can be used with different queries. However, available optimization techniques
can be used to specialize the program for a given query (cf.  \cite{monica10,bertossi11} for this kind of optimizations for
repair logic programs).

\vspace{-2mm}
\section{The CQA Connection}\label{sec:cqa}
Consider a database instance $D$ that fails to satisfy a given set of integrity constraints $\nit{IC}$. It still contains useful and some semantically correct information. The area of {\em consistent query answering} (CQA) \cite{bertossi06,bertossi11} has to do with: (a)
Characterizing the information in $D$ that is still semantically correct wrt $\nit{IC}$, and (b) Characterizing, and
computing, in particular, the semantically correct, i.e. consistent, answers to a query $\mc{Q}$ from $D$ wrt $\nit{IC}$. The first goal is achieved by proposing a {\em repair semantics}, i.e. a class of alternative instances
to $D$ that are consistent wrt $\nit{IC}$ and minimally depart from $D$. The consistent information in $D$ is the one that is invariant under all the repairs in the class. This applies in particular to the consistent answers: They should hold in every minimally repaired instance.

There are some connections between CQA and our treatment of privacy preserving query answering. Notice that
every view definition of the form (\ref{eq:view}) can be seen as an integrity constraint expressed in
the FO language $L(\Sigma \cup \{V_s\})$:
\begin{equation}
\forall \bar{x}(V_s({\bar x}) \ \longleftrightarrow \  \exists \bar{y}(R_1(\bar{x}_1)
\wedge \cdots \wedge R_n(\bar{x}_n) \wedge \varphi)), \label{eq:icview}
\end{equation}
with $\bar{y} = (\bigcup \bar{x}_i) \smallsetminus \bar{x}$. From this perspective, the problem of {\em view maintenance}, i.e. of maintaining the view defined by (\ref{eq:icview})  synchronized with the base relations \cite{gupta95}  becomes a problem of {\em database maintenance}, i.e. maintenance of the consistency of the
database wrt (\ref{eq:icview}) seen as an IC. This also works in the other direction since every IC can be associated to a violation view, which
has to stay empty for the IC to stay satisfied.

Actually, we want more than maintaining the view defined in (\ref{eq:icview}). We want it to be empty or returning
only tuples with null values. In consequence, we have to impose the following ICs on $D$, which are obtained from
the RHS of (\ref{eq:icview}): If $\bar{x}$  is $x^1, \ldots, x^k$, then for $1 \leq i \leq k$,
\begin{equation}
\forall \bar{x} \bar{y} \neg (R_1(\bar{x}_1)
\wedge \cdots \wedge R_n(\bar{x}_n) \wedge \varphi \wedge x^i \neq \nn). \label{eq:ics}
\end{equation}
That is, from each view definition (\ref{eq:icview}) we obtain $k$ {\em denial constraints} (DCs), i.e. prohibited
conjunctions of (positive) database atoms and built-ins. DCs have been investigated in CQA  under
several repair semantics \cite{chM05,bertossi11}.

In our case,
the secrecy instances correspond to the repairs of $D$ wrt the set DCs  in (\ref{eq:ics}). These repairs
are defined according to the null-based (and attribute-based \cite{bertossi11}) repair semantics  of Section \ref{sec:secInstance}, i.e.
$\leq_D$-minimality (cf. Example \ref{ex:minimal}).
Through this correspondence we can benefit from concepts and techniques developed for CQA.
\begin{example}
The secrecy view defined by

\vspace{1mm}
\centerline{$\Vs(x, z)\leftarrow P(x,y),R(y,z),y<3$}

\vspace{1mm}\noindent  gives rise to the following denial
constraints: \
$\neg \exists xyz(P(x,y) \wedge R(y,z) \wedge y<3 \wedge x \neq \nn)$ and
$\neg \exists xyz(P(x,y) \wedge R(y,z) \wedge y<3 \wedge z \neq \nn)$. \
A instance $D$ has to be minimally repaired in order to satisfy them. \boxtheorem
\end{example}

\vspace{-3mm}
\section{Related Work}\label{sec:related}
Other researchers have investigated the problem of data privacy and access control in relational databases.
We described in Section \ref{sec:intro} the approach based on authorization views \cite{RMS04,ZM05}. In \cite{LA04},
the privacy is specified through values in cells within tables that can be accessed by a user.
To answer a query $\mc{Q}$ without violating privacy, they propose the table and query semantics models, which
generate masked versions of the tables by replacing all the cells that are not allowed to be accessed with {\tt NULL}.
When the user issues $\mc{Q}$, the latter is posed to the masked versions of the tables, and  answered as usual.
The table  semantics is independent of any queries, and views. However, the query semantics takes queries into account.
\cite{LA04} shows the implementation of two models based on query rewriting.

Recent work \cite{wang07} has presented a labeling approach for masking unauthorized information by using two types of
 special variables. They propose a secure and sound query evaluation algorithm in the case of
 cell-level disclosure policies, which determine for each cell whether the cell is allowed to be accessed or not. The
 algorithm is based on query modification, into one that returns less information than the original one. Those approaches
 propose query rewiring to enforce fine-grained access control in databases. Their approach is mainly algorithmic.

Data privacy and access control in incomplete propositional databases has been studied
in \cite{bw07,bw08,weibert1}. They take a different approach, {\em control query evaluation} (CQE), to fine-grained access
control. It is policy-driven, and aims to ensure confidentiality on the
basis of a logical framework. A security policy specifies the facts that a certain user is not allowed to access. Each
query posed to the database by that user is checked, as to  whether the answers to it would allow the user to infer any
sensitive information. If that is the case, the answer is distorted by either {\em lying} or {\em refusal} or
{\em combined lying and refusal}. In \cite{btw10}, they extend CQE to restricted incomplete FO logic databases
via a  transformation  into a propositional language.
This approach seem to be incomparable to ours. They do not use null values, and the issue of maximality of answers
that do not compromise privacy is not explicitly addressed.

Our approach is based on producing virtual updates on the database, by forcing the secrecy views to become null.
This is clearly reminiscent of the older, but still challenging  database problem of updating a
database  through views \cite{cosmadakis}. Here we confront new difficulties, namely the occurrence of SQL nulls with a
special semantics, and the minimality of null-based changes on the base relations.

In \cite{bb06} a null-based repair semantics was introduced, but it differs from the one introduced in Section
\ref{sec:secInstance}. The former was proposed for enforcing satisfaction of sets of ICs that include
referential ICs, which require the possible insertion of new tuples with nulls. The comparison between instances
is based onsets of full tuples and also on the occurrence of nulls in them. Here, we enforce secrecy by changes of
attributes values only.

A representation of null values in logic programs with stable model semantics is proposed in \cite{gelfond94}, whose
aim is to capture the intended semantics of null values {\em \`{a} la} Reiter, i.e. as found in his logical reconstruction
of relational databases \cite{reiter84}. Two remarks have to be made here. First, Reiter reconstructs ``logical"
nulls, but not SQL nulls. In our work we use the latter, as done in database practice. Second, we take care of nulls by
proposing a new query answering semantics that can be captured in classic logical terms via query rewriting. The rewritten
queries are the input to a logic program, which then treats them as ordinary constants (without having to give a logical
account of them).

\vspace{-2mm}
\section{Conclusions}\label{sec:conclusions}
In this work, we have developed a logical framework and a methodology for answering conjunctive queries that do not reveal
secret information as specified by secrecy views. Our work is of a foundational nature, and attempts to provide a
theoretical basis, or at least part of that basis, for possible technological developments. Implementation efforts
and experiments, beyond the proof-of-concept examples we have run with {\em DLV}, are left for future work.

We have concentrated on conjunctive secrecy views and conjunctive queries. We have assumed that the databases may
contain nulls, and also nulls are used to protect secret information, by virtually updating with nulls some of the
attribute values. In each of the resulting alternative virtual instances, the secrecy views either become
empty or contain a tuple showing only null values. The queries can be posed against any of these virtual
instances or cautiously against all of them, simultaneously. The latter guarantees privacy.

The update semantics enforces (or captures) two natural requirements. That the updates are based on  null values, and
that the updated instances stay close to the given instance. In this way, the query answers become implicitly
maximally informative, while not revealing the original contents of the secrecy views.

The null values are treated as in the SQL standard, which in our case, and for conjunctive query answering,  is
reconstructed in classical logic. This reconstruction captures well the ``semantics" of SQL nulls (which in not clear
or complete in the standard), at least for the case of conjunctive query answering, and some extensions thereof. This
is the main reason for concentrating on conjunctive queries and views. In this case, queries and views can be
syntactically transformed into conjunctive queries and views for which the evaluation or verification can be done by
treating nulls  as any other constant.

The secret answers are based on a skeptical semantics. In principle, we could consider instead the more relaxed
{\em possible} or {\em brave}
semantics: an answer would be  returned if it holds {\em in some} of the secrecy instances. The
{\em possibly secret answers} would provide more information about the original database than the (certainly) secret answers.
However, they are not suitable for our the privacy problem.
\begin{example} (example  \ref{ex:minimal} continued)   A {\em possibly
secret answer} to the query $\mc{Q}_1(x,y): P(x,y)$ is $\langle 1,2\rangle$, obtained from $D_3$. Similarly, $\langle 2,1\rangle$
is a possibly secret answer to $\mc{Q}_2(x,y): R(x,y)$. From these possibly secret
answers, the user can obtain the contents of the secrecy view. \boxtheorem
\end{example}
We introduced disjunctive logic programs with stable model semantics  to specify the secrecy instances.  This is a single
program that can be used to compute secret answers to any conjunctive query. This provides a general mechanism, but may
not be the most efficient way to go for some classes of
secrecy views and queries. {\em Ad hoc} methods could be proposed for them, as has been the case in CQA \cite{amw09,bertossi11}.

Our work leaves several open problems, and they are matter of ongoing and future research. Complexity issues have
to be explored. For example, of deciding whether or not a particular instance is a secrecy instance of an original instance.
Also, of deciding if a tuple is a secret answer to a query. The connection with CQA, where similar
problems have been investigated, looks very promising in this regard.

Another problem is about query rewriting, i.e. about the possibility of rewriting the original query
into a new FO query, in such a way that the new query, when answered by the given instance, returns the
secret answers. From the connection with CQA we can predict that this approach has limited applicability,
but whenever possible, it should be used, for its simplicity and lower complexity.

For future work, it would be interesting to investigate the connections with {\em view determinacy} \cite{nash10},
that has to do with the possible determination of extensions of query answers by a set of views with a fixed
contents. The occurrence of SQL nulls and their semantics introduces a completely new dimension into this problem.

A natural extension of this work would go in the direction of freeing ourselves from the assumptions listed at the
end of Section \ref{sec:secretAns}.  Their relaxation would create a  challenging new scenario, and most likely, would
require a non-straightforward modification of our approach. One of these possible relaxations consists in the addition of
ICs to the schema. If they are known to the user, and, most importantly, that they are satisfied by the database, then
privacy could be compromised. Also the updates leading to the virtual updates should take these ICs into account, to
produce consistent secrecy instances.

It would also be interesting to investigate  more expressive queries and secrecy views, going beyond the
conjunctive case. However, if we allow negation, the challenges become intrinsically more difficult.
 On one side, in the case of secrecy views, negation becomes  a fundamental complication for privacy
\cite{RMS04,ZM05}. On the other, the query rewriting methodology that captures nulls as ordinary constants (cf. Section
\ref{app:nans}) that we have used in our work does not include the combination of nulls and negation. The extension of
our privacy approach to queries or secrecy views with negation would make it  necessary to first attempt an extension of
this kind of query rewriting. However, this requires to agree on a sensible semantics for SQL nulls in the context
of such more expressive queries, something that is definitely worth investigating.

\vspace{2mm}\noindent
{\small {\bf Acknowledgements:} \ This research started when Leo Bertossi was spending his sabbatical at the TU
Vienna. Support from Georg Gottlob, Thomas Eiter and a Pauli Fellowship  of the
``Wolfgang Pauli Institute, Vienna" is highly appreciated.
We are indebted to Thomas Eiter and Loreto Bravo for technical conversations at an early stage of this research,
and to Sina Ariyan for some computational experiments.  Research funded by NSERC Discovery and
NSERC/IBM CRDPJ/371084-2008.}

\bibliographystyle{plain}

\ignore{

}

\vspace{2mm}
\noindent {\bf Leopoldo Bertossi}  has been Full Professor at the School of
Computer Science, Carleton University (Ottawa, Canada) since
2001. He is Faculty Fellow of the IBM Center for Advanced
Studies. He obtained a PhD in Mathematics
from the Pontifical Catholic University of Chile (PUC) in 1988.
Until 2001 he was professor
at the Department of Computer Science, PUC; and also the
President of the Chilean Computer Science Society (SCCC) in
1996 and 1999-2000.
\ignore{He has held visiting positions at the computer science
departments of the universities of Toronto (1989/90),
Wisconsin-Milwaukee (1990/91), Marseille-Luminy (1997), Technical
University Berlin (1997/98), Free University of Bolzano-Bozen (2005),
and the Technical University of Vienna as a Pauli
Fellow of the ``Wolfgang Pauli Institute (WPI) Vienna".}
His research interests include database theory,
data integration, peer data management,
intelligent information systems, data quality,
knowledge representation, and answer set programming.

\vspace{2mm}
\noindent {\bf Lechen Li} was born in Sichuan, China in 1985. She received a Bachelor in Computer Engineering
from the Sichuan Normal University, Chengdu, China, and a MSc degree in computer science in 2011 from Carleton University, Ottawa, Canada, under the
supervision of Prof. L. Bertossi. Her master's research was in the area of data privacy.

\end{document}